\documentclass[12pt]{article}
\usepackage{latexsym,amsmath,theorem,amssymb,biometri,graphicx,mathrsfs,
  setspace}

\let\bibhang\relax
\usepackage{natbib}

\usepackage{url}

\usepackage[ruled,vlined]{algorithm2e}

\usepackage{booktabs}

\usepackage{rotating}

\allowdisplaybreaks

\DeclareMathOperator{\sign}{sign}

\newcommand{\nc}{\newcommand}

\newcommand*\xbar[1]{%
  \hbox{%
    \vbox{%
      \hrule height 0.5pt 
      \kern0.275ex
      \hbox{%
        \kern0em
        \ensuremath{#1}%
        \kern0em
      }%
    }%
  }%
} 

\nc{\w}{\wedge}
\nc{\h}{\widehat}
\nc{\td}{\widetilde}
\nc{\T}{\top}

\nc{\e}{E}
\nc{\he}{\h{\e}}

\nc{\al}{\pmb{\eta}}
\nc{\ha}{\h{\al}}
\nc{\ba}{\xbar{\al}}

\nc{\ta}{\td{\al}}

\nc{\bt}{\pmb{\beta}}
\nc{\hb}{\h{\bt}}
\nc{\tb}{\td{\bt}}

\DeclareFontFamily{U}{mathx}{}
\DeclareFontShape{U}{mathx}{m}{n}{<-> mathx10}{}
\DeclareSymbolFont{mathx}{U}{mathx}{m}{n}
\DeclareMathAccent{\widecheck}{0}{mathx}{"71}
\nc{\cb}{\widecheck{\bt}}

\nc{\epi}{\varepsilon}
\nc{\dconv}{\stackrel{d}{\rightarrow}}

\let\S\relax
\nc{\S}{Section}

\theoremstyle{bkaasp}
\newtheorem{condition}{Condition}
\theoremstyle{bkathm}

\newtheorem{theorem}{Theorem}
\newtheorem{corollary}{Corollary}
\newtheorem{prop}{Proposition}

\newtheorem{remark}{Remark}
\newtheorem{example}{Example}

\begin{document}
\baselineskip=25pt

\begin{center}
\hfill{\Large\bf Cross-Audit Projection for Model Risk Prediction}\\
      {\large\sc Yijian Huang}\\
      Department of Biostatistics and Bioinformatics,
  Emory University,\\
  Atlanta, Georgia 30322, U.S.A.\\
  yhuang5@emory.edu
\end{center}
\vspace*{.1in}

\section*{Abstract}
For training-data-based model risk prediction, $K$-fold cross-validation~(CV)
is widely used to mitigate the well-known over-optimism of the empirical risk
and is often regarded as reliable. However, for binary
classification via empirical risk minimization, our numerical studies
reveal a surprising phenomenon: $K$-fold CV may perform poorly in estimating
class-specific risks, even worse than the empirical estimator. We
perform a higher-order asymptotic analysis showing that $K$-fold CV may
converge at a slower rate, whereas the empirical estimator exhibits a
second-order asymptotic bias that explains its over-optimism. These findings
motivate a novel two-step procedure for model risk prediction, termed
cross-audit projection (CAP). The cross-audit step adopts the same resampling
scheme as $K$-fold CV to estimate over-optimism in subsamples, while the
asymptotic-theory-informed projection step adjusts for the reduced sample size
in bias correction of the
empirical risk. The resulting CAP estimator is first-order
asymptotically equivalent to the empirical risk while achieving second-order
asymptotic unbiasedness. An accompanying inference procedure is also developed.
Simulation studies support theoretical advantages of CAP and
demonstrate favorable finite-sample performance.
An application to breast cancer detection further illustrates the proposed
method.

\noindent KEY WORDS:
Asymptotic bias;
Bias correction;
Binary classification;
Cross-validation;
Empirical risk minimization;
Over-optimism.

\vspace*{.3in}

\section{Introduction}

Once a predictive model is trained, its performance in future applications
needs to be evaluated to inform potential deployment. The empirical risk
is well known to be optimistically biased. Although independent
validation is conceptually natural, reserving data for this
purpose leads to inefficient use of typically limited samples.
Accurate model risk prediction based on the training data has long been pursued.

Classical approaches seek to correct empirical risk analytically, as in
Mallows’ $C_p$ \citep{mallows}, AIC \citep{akaike}, and covariance penalties
\citep{stein,efron04}. However, these methods are model-based, restrictive in
their choice of loss function, and are not well suited to increasingly complex
modern predictive algorithms. Resampling-based
methods provide a natural alternative to overcome these limitations.
\citet{harrell} corrected empirical risk using the
nonparametric bootstrap for bias estimation. Nevertheless, the bias may
not be eliminated because bootstrap
resamples overlap with the original data.
Cross-validation (CV) methods \citep{allen,geisser,stone74,stone77}, which
keeps training and validation subsamples mutually exclusive, have
become the predominant tools. Depending on how training
subsamples are constructed, variants include leave-one-out CV, $K$-fold CV
with a typical choice of $K=10$, repeated $K$-fold CV, repeated
learning-testing schemes \citep{burman}, and bootstrap-based methods such as
the 0.632 and 0.632+ rules \citep{efron83,efron97}.

Among CV methods, $K$-fold CV is perhaps the most widely used owing to its
conceptual simplicity and moderate computational cost; in contrast,
leave-one-out CV can be computationally prohibitive for modern learning
algorithms. However, we reveal a previously undocumented phenomenon that
$K$-fold CV may perform poorly despite its perceived reliability, through
an example of binary classification via empirical risk minimization (ERM).

\begin{figure}[t]
  \centerline{\includegraphics[width=6.5in]{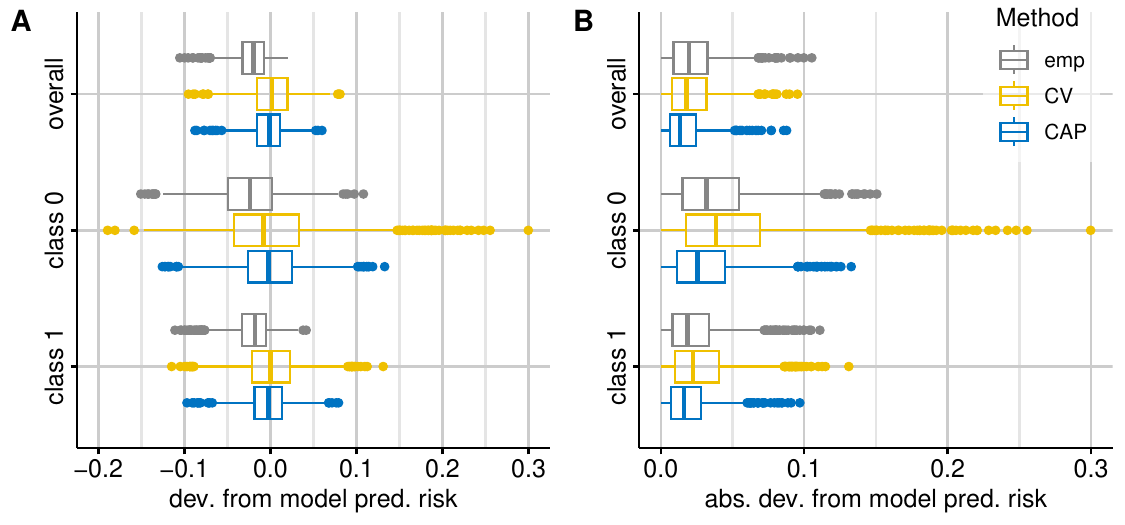}}
          {\caption{\label{fig1} Simulation results on empirical, 10-fold
              CV, and proposed CAP estimators of overall and class-specific
              model prediction risks in single-feature
          binary ERM classification with
          $X_0\sim N(0,1)$, $X_1\sim N(1,1)$, $\omega=0.25$, and
          $n_0=n_1=100$, based on 1000 simulated datasets. Panel A: Boxplots of
          deviations from model prediction risks. Panel B: Boxplots of
          absolute deviations from model prediction risks.}}
\end{figure}

\begin{example}[ERM classifier with a single feature]\label{expl1}\rm
  This binary classification is based on the dichotomization of a single
  feature, denoted by $X_d$ for class $d \in \{0,1\}$. To accommodate
  differential costs of class-specific misclassification,
  the overall risk is defined as a weighted sum of class-specific
  misclassification rates, with weights $\omega$
  and $1-\omega$ for classes~0 and~1, respectively.
  A classifier that minimizes the overall risk is considered optimal,
  and ERM is employed for the estimation. Suppose that $n_d$ independent
  replicates of $X_d$ are observed for each class $d=0,1$. Figure~\ref{fig1}
  presents simulation results of the empirical risks and 10-fold CV estimates.
  As expected, CV mitigated over-optimism of the empirical risks, overall
  and class-specific alike. However, surprisingly CV tracked the
  class-specific risks even less accurately than the empirical estimator.
  \label{exp1}
\end{example}

Through internal validation, $K$-fold CV directly and unbiasedly targets the
average risk of the $K$ subsample models, known as the
{\em CV ensemble prediction risk}. Its discrepancy from the model prediction
risk is commonly assumed to be negligible. The same
premise also underlies several inferential procedures. \citet{dudoit},
\citet{austern}, and
\citet{bayle} established central limit theorems for $K$-fold CV with
respect to the CV ensemble prediction risk for various learning algorithms,
and the resulting methods have been used to draw inference on the model
prediction risk; see also \citet{ledell} and \citet{benkeser}. \citet{bates}
proposed an additional inference procedure based on $K$-fold CV. However,
in light of the unexpected phenomenon observed in Example~\ref{expl1},
a closer examination of $K$-fold CV is warranted, along with a broader
question of whether alternative methods can improve the estimation and inference
of model prediction risk. To address these issues, we will consider general linear ERM
classification and conduct an asymptotic study to elucidate the anomalous behavior of $K$-fold CV, as well as
the over-optimism of empirical risk. These findings
motivate a novel method termed cross-audit projection (CAP). For a
preview, Figure~\ref{fig1} also presents the CAP estimates, illustrating their
superior performance.

The contributions of this article are four-fold. First, to our knowledge, we
demonstrate for the first time that $K$-fold CV may perform worse than
the empirical risk. Second, we establish a higher-order asymptotic
analysis framework to characterize both bias and variability in model risk
prediction. This framework clarifies the over-optimism in the empirical
risk through second-order asymptotic bias and reveals limitations of
$K$-fold CV. Third, and most importantly, we develop the CAP method for accurate
model risk prediction by devising an asymptotic-theory-informed resampling
procedure. Finally, we provide a refined treatment of ERM for binary
classification that is of independent interest.
The remainder of the article is organized as follows. Section~2 formulates
linear ERM classifier and establishes its asymptotic properties. Section~3
investigates the asymptotic behaviors of
empirical and $K$-fold CV risk estimators, which in turn motivate the CAP
procedure for point prediction and inference. Simulation studies and an
application to
breast cancer detection are presented in Section~4. Section~5 concludes with
a discussion. Technical
proofs are deferred to the Appendix, with additional results provided in the
Supplementary Materials.

\section{Linear ERM classification}

Consider binary classification with an $m$-vector feature, ${\bf X}_d$, for
$m\geq1$ and class~$d\in\{0,1\}$. A linear classifier indexed by $(m+1)$-vector
coefficient $\bf b$ predicts the class label as
$I({\bf b}^\T{\bf X}_d^\circ>0)$, where
${\bf X}_d^\circ\equiv({\bf X}_d^\T,1)^\T$; the last component of $\bf b$
represents minus threshold for the feature combination.
The two class-specific risks and overall risk are given by
\[
  \psi_0({\bf b})=\Pr({\bf b}^\T{\bf X}_0^\circ>0),\quad
\psi_1({\bf b})=\Pr({\bf b}^\T{\bf X}_1^\circ\leq0),\qquad
\psi({\bf b})=\omega \psi_0({\bf b})+(1-\omega)\psi_1({\bf b});
\]
recall that $\omega\in(0,1)$ is a user-specified relative weight between
the two classes. Two classifiers are said to be
{\em parallel} if their coefficients differ by a scaling factor; such
classifiers have identical risks. The optimal classifier minimizes
the overall risk, subject to a norm constraint for
identifiability:
\[
  \bt=\arg\min_{{\bf b}:\|{\bf b}\|_1=1}\psi({\bf b}),
\]
where $\|\cdot\|_1$ denotes the $\ell_1$ norm.
The choice of $\ell_1$ norm will facilitate computation in classifier
learning but is otherwise non-essential.

\subsection{Learning via ERM}

Our ERM learning method is closely related to that of \citet{elliott} under
the cohort design, where the celebrated maximum score estimator of
\cite{manski75,manski85} is adapted to accommodate class-specific
utilities. We adopt the case-control design instead, with $n_d$ independent
replicates of
${\bf X}_d$ observed: ${\bf X}_{d,[i]}$, $i=1,\ldots,n_d$, for $d=0,1$. Write
$n\equiv n_0+n_1$. The
class-specific empirical risks and overall empirical risk are
\[ \h{\psi}_0({\bf b})=\he I({\bf b}^\T{\bf X}_0^\circ>0),\quad
\h{\psi}_1({\bf b})=\he I({\bf b}^\T{\bf X}_1^\circ\leq0),\qquad
\h{\psi}({\bf b})=\omega\h{\psi}_0({\bf b})+(1-\omega)\h{\psi}_1({\bf b}),
\]
where $\he$ represents empirical average, e.g.,
$\he I({\bf b}^\T{\bf X}_0^\circ>0)=
n_0^{-1}\sum_{i=1}^{n_0}I({\bf b}^\T{\bf X}_{0,[i]}^\circ>0)$.
ERM corresponds to the following optimization problem,
\begin{equation}
  \min_{{\bf b}:\|{\bf b}\|_1=1}\h{\psi}({\bf b}),
  \label{erm}
\end{equation}
and the estimated classifier $\hb$ is a near minimizer, satisfying
\begin{equation}
  \h{\psi}(\hb)\leq \min_{{\bf b}:\|{\bf b}\|_1=1}\h{\psi}({\bf b})+\epi_n,
  \quad \epi_n=o(n^{-2/3}).
  \label{est}
\end{equation}
The classifier in Example~\ref{exp1} is a special case with $m=1$.

The computation of problem~(\ref{erm}) is challenging due to the non-convex,
piecewise-constant objective function and the presence of an equality
constraint. \cite{huang22} proposed a novel and effective algorithm tailored
to this class of problems.
The equality constraint is handled through
a reformulation by incorporating a penalty term:
\begin{equation}
  \min_{{\bf b}:\|{\bf b}\|_1\leq1}\h{\psi}({\bf b})-c\|{\bf b}\|_1,
  \label{est2}
\end{equation}
for a constant $c>0$. To address the non-smoothness, indicator
function $I(x\leq0)$ is approximated using
$\sigma^{-1}\{(x-\sigma/2)^--(x+\sigma/2)^-\}$, where $x^-\equiv-\min(x,0)$
and $\sigma>0$ is a smoothing parameter. This approximation also leads to a
concave-convex decomposition of the objective function, enabling an
application of the concave-convex procedure \citep{yuille}.
Note that taking $c>1$ in the smoothed problem ensures its minimizer
bounded away from 0; see related
discussion in \citet[][section~4]{huang22}.
The algorithm proceeds by solving a sequence of smoothed problems that
converge to the original one, with decreasing $\sigma$ values. More details
are provided in Supplementary Appendix~A.

\subsection{Asymptotic theory on the ERM classifier}

The ERM classifier is studied to provide a foundation for subsequent
analyses of model risk prediction methods. Existing results for related
estimators are extended and generalized. For the maximum
score estimator, consistency \citep{manski75,manski85} and weak convergence
with cube-root asymptotics \citep{kim} were established under a
semiparametric quantile regression model. \citet{elliott} considered a
nonparametric model similar to ours, but established only consistency for
their estimator.

Focus on consistency first, with the following regularity conditions.
\begin{condition}[Sample sizes]\label{con1}
  As $n\rightarrow\infty$, $n_1/n_0$ converges to a finite constant $\gamma>0$.
\end{condition}
\begin{condition}[Identifiability and separation]\label{con2}
  The overall risk $\psi({\bf b})$ and optimal classifier $\bt$ satisfy
  $\psi(\bt)<\inf_{{\bf b}:\|{\bf b}\|=1,\|{\bf b}-\bt\|_1\geq\epi}\psi({\bf b})$
  for every $\epi>0$.
\end{condition}

\begin{theorem}\label{thm1}
  Under Conditions~\ref{con1}--\ref{con2}, $\hb$ converges to
  $\bt$ almost surely, and $\h{\psi}(\hb)$ is strongly consistent for
  $\psi(\hb)$.
\end{theorem}
This consistency result actually holds more generally, allowing $\epi_n=o(1)$
instead in the definition~(\ref{est}) of $\hb$.

Now, we establish weak convergence in the cube-root
asymptotics framework of \cite{kim}. Several additional technical
conditions are imposed. For a generic vector $\bf b$, write ${\bf b}_\ell$
as the $\ell$-th element and ${\bf b}_{-\ell}$ as the vector with the $\ell$-th
element removed.
\begin{condition}[Distribution smoothness]\label{con3}
  Two cases are considered based on the number of features: For $d=0,1$,
  \begin{list}{}{\itemsep 0ex\parsep 0ex\topsep 0ex}
  \item[(A) $m=1$:] the density function of $\bt^\T{\bf X}_d^\circ$
    at 0 exists and is bounded;
  \item[(B) $m\geq2$:] there exists $\ell$, $1\leq\ell\leq m$,
    such that (i)~$\bt_\ell\neq0$, (ii)~${\bf X}_{d,-\ell}$ is integrable, and
    (iii)~the conditional density function of
    $\bt^\T{\bf X}_d^\circ\mid {\bf X}_{d,-\ell}$ exists and is bounded.
  \end{list}
\end{condition}
Since the classification is invariant under linear transformation of the
features, Case~(B) may be interpreted as corresponding to appropriately
transformed features. As such, it is more general than it may initially appear.
\begin{condition}[Hessian of overall risk]\label{con4}
  The Hessian $\nabla^2\psi({\bf b})$ exists at ${\bf b}=\bt$, and satisfies
  ${\bf h}^\T\nabla^2\psi(\bt){\bf h}>0$ for ${\bf h}\neq{\bf 0}$ in the
  subspace
  orthogonal to $\bt$.
\end{condition}
\begin{condition}[Nondegeneracy of limiting process]\label{con5}
  For ${\bf h}\neq{\bf 0}$ in the subspace orthogonal to $\bt$,
  ${\bf h}^\T\nabla\psi_d(\bt)\neq0$ for $d=0,1$.
\end{condition}
Restricting $\bf h$ to the subspace in
Conditions~\ref{con4}--\ref{con5} is linked to the scale invariance of
the risks. These conditions are otherwise standard in $M$-estimation.

The limiting distribution of $\hb$ is degenerate by definition.
To avoid the complication, we focus on a related theoretical
classifier instead, having its $\ell$-th coefficient fixed to $\bt_\ell$:
\[
\cb\equiv
\left\{\begin{array}{lll}\displaystyle
\hb\bt_\ell/\h{\bt}_\ell
&\hspace*{.2in}& \bt_\ell\h{\bt}_\ell>0\\[8pt]
\mbox{$\bt$} &&\mathrm{otherwise}
\end{array}\right..
\]
Since $\hb$ is consistent for $\bt$, so is $\cb$. Furthermore, with
probability tending to 1, $\cb$ is parallel to $\hb$.
Write $\phi_d({\bf b}_{-\ell})\equiv\psi_d({\bf b})|_{{\bf b}_\ell=\bt_\ell}$, $d=0,1$,
$\phi({\bf b}_{-\ell})\equiv\psi({\bf b})|_{{\bf b}_\ell=\bt_\ell}$, and their empirical
counterparts as $\h{\phi}_d({\bf b}_{-\ell})$ and $\h{\phi}({\bf b}_{-\ell})$,
respectively.
For asymptotic analysis, $\cb_{-\ell}$ can be characterized as a near minimizer
of an unconstrained problem:
\[
\h{\phi}(\cb_{-\ell})\leq\min_{{\bf b}_{-\ell}}\h{\phi}({\bf b}_{-\ell})+o_p(n^{-2/3}).
\]
However, $\cb$ is not a practical classifier since the estimand $\bt$ is
involved.

Denote weak convergence by $\rightsquigarrow$.
\begin{theorem}\label{thm2}
  Suppose that Conditions~\ref{con1}--\ref{con5} hold. Then,
  \begin{equation}
  n^{2/3}\left\{\begin{array}{c}\displaystyle
    \h{\phi}_0(\bt_{-\ell}+n^{-1/3}{\bf s})-\phi_0(\bt_{-\ell}+n^{-1/3}{\bf s})
    -\h{\phi}_0(\bt_{-\ell})+\phi_0(\bt_{-\ell})\\
    \h{\phi}_1(\bt_{-\ell}+n^{-1/3}{\bf s})-\phi_1(\bt_{-\ell}+n^{-1/3}{\bf s})
    -\h{\phi}_1(\bt_{-\ell})+\phi_1(\bt_{-\ell})\\
    \h{\phi}(\bt_{-\ell}+n^{-1/3}{\bf s})-\h{\phi}(\bt_{-\ell})
  \end{array}\right\}\rightsquigarrow
  \left\{\begin{array}{c}\displaystyle
  W_0({\bf s})\\
  W_1({\bf s})\\
  Z({\bf s})\end{array}\right\},\label{wc}
  \end{equation}
  where $W_d({\bf s})$, $d=0,1$, are two independent mean-zero
  Gaussian processes with continuous sample paths, and
  $Z({\bf s})=W({\bf s})+{\bf s}^\T{\bf H}{\bf s}/2$ for
  $W({\bf s})=\omega W_0({\bf s})+(1-\omega)W_1({\bf s})$ and
  ${\bf H}=\nabla^2\phi(\bt_{-\ell})$. Furthermore,
  $n^{1/3}(\cb_{-\ell}-\bt_{-\ell})\rightsquigarrow
  {\bf U}\equiv\arg\min_{\bf s} Z({\bf s})$.
\end{theorem}

Weak convergence of $\hb$ follows immediately. Nevertheless, the results on
$\cb$ suffice for the risk prediction analysis of the
estimated classifier, which is the primary focus of this article.

\section{Model risk prediction}

Over-optimism in empirical risk can be {\em exactly} characterized
for simple models and certain loss functions, as exploited in classical
correction methods \cite[e.g.,][]{efron04}. However, such characterization
becomes challenging for more complex predictive algorithms, including
linear ERM classification. On the other hand, $K$-fold CV has rarely been
analyzed with respect to the model prediction risk.
A novel theoretical framework for assessing model risk prediction methods is
therefore needed, which we develop based on higher-order asymptotics.
We first elucidate the over-optimism in the empirical risk
and the limitations of $K$-fold CV, which in turn motivate
our proposed CAP procedure.

\subsection{Asymptotic theory on empirical and $K$-fold CV risks}

With estimated classifier $\hb$, the overall model prediction risk is given
by $\psi(\hb)$ and class-specific ones by $\psi_d(\hb)$,
$d=0,1$. Building on the weak convergence results in Section~2.2, we obtain
second-order expansions of the empirical risks to characterize the over-optimism
through asymptotic bias \citep[e.g.,][section~2.5.2]{shao}.
\begin{theorem}\label{thm3}
  Under Conditions~\ref{con1}--\ref{con5},
  \begin{eqnarray}
    n^{2/3}[\{\h{\psi}_d(\hb)-\psi_d(\hb)\} - \{\h{\psi}_d(\bt)-\psi_d(\bt)\}]
    &\rightsquigarrow& W_d({\bf U}),\qquad d=0,1,\label{opt1}\\
    n^{2/3}[\{\h{\psi}(\hb)-\psi(\hb)\} - \{\h{\psi}(\bt)-\psi(\bt)\}]
    &\rightsquigarrow& W({\bf U})=Z({\bf U})-{\bf U}^\T{\bf HU}/2,\label{opt2}
  \end{eqnarray}
  where $W_d({\bf U})$, $d=0,1$, have negative expectations, and
  $W({\bf U})$ is negative almost surely.
\end{theorem}
Since $E\h{\psi}_d(\bt)=\psi_d(\bt)$ and $E\h{\psi}(\bt)=\psi(\bt)$, the
empirical risks $\h{\psi}_d(\hb)$ and $\h{\psi}(\hb)$ are first-order
asymptotically unbiased for the corresponding prediction risks $\psi_d(\hb)$
and $\psi(\hb)$. Similar first-order asymptotic results have been
established previously \citep[e.g.,][]{dudoit}, but they do not account for
the observed poor performance of empirical risks. As a key and novel feature
of our framework, the second-order asymptotic bias explains and characterizes
the over-optimism. Although the over-optimism can be explained for the overall
risk without resorting to asymptotics, this explanation is qualitative and
does not extend to the class-specific risks.

With $K$-fold CV, the training data are randomly
partitioned into $K$ roughly
equal-sized and non-overlapping folds, stratified by class. For each
$k=1,\ldots,K$, let $\hb^{(-k)}$ be the counterpart of
$\hb$ using the data except the $k$-th fold, $\h{\psi}_d^{(k)}({\bf b})$ be the
counterpart of $\h{\psi}_d({\bf b})$ using the $k$-th fold only, for $d=0,1$. The
CV class-specific and overall risks are given by
\begin{equation}
\xbar{\psi}_{d,\mathrm{cv}}=K^{-1}\sum_{k=1}^K\h{\psi}_d^{(k)}(\hb^{(-k)}),\quad d=0,1,
\qquad \xbar{\psi}_{\mathrm{cv}}=\omega\xbar{\psi}_{0,\mathrm{cv}}
+(1-\omega)\xbar{\psi}_{1,\mathrm{cv}},
\label{cvrisk}
\end{equation}
respectively. Because of random splitting, $\hb^{(-k)}$ is independent of
$\h{\psi}_d^{(k)}({\bf b})$. Thus, the CV risks are unbiased estimators of
the CV ensemble prediction risks discussed in Section~1,
\[
\psi_{d,cv-ens}\equiv K^{-1}\sum_{k=1}^K\psi_d(\hb^{(-k)}),\quad d=0,1,
\qquad \psi_{cv-ens}\equiv K^{-1}\sum_{k=1}^K\psi(\hb^{(-k)}),
\]
respectively.
\begin{corollary}\label{coro2}
  Suppose that $K\geq2$ is fixed. Under Conditions~\ref{con1}--\ref{con5},
  \begin{eqnarray}
    n^{2/3}\left[(\xbar{\psi}_{d,\mathrm{cv}}-\psi_{d,cv-ens}) - \{\h{\psi}_d(\bt)-\psi_d(\bt)\}\right]
    &\rightsquigarrow& V_d,\qquad d=0,1,\label{cv1}\\
    n^{2/3}\left[(\xbar{\psi}_{\mathrm{cv}}-\psi_{cv-ens}) - \{\h{\psi}(\bt)-\psi(\bt)\}\right]
    &\rightsquigarrow& V,\label{cv2}\\
    n^{1/3}\left\{\psi_{d,cv-ens}-\psi_d(\hb)\right\}
    &\rightsquigarrow& S_d,\qquad d=0,1,\label{cvi}\\
    n^{2/3}\left\{\psi_{cv-ens}-\psi(\hb)\right\}
    &\rightsquigarrow& S,\label{cvii}
  \end{eqnarray}
  for some non-degenerate random variables $V_d$, $V$, $S_d$, and $S$, where
  $V_d$, $V$, and $S_d$ have zero means and
  $ES=[\{(K-1)/K\}^{-2/3}-1]E{\bf U}^\T{\bf HU}/2>0$.
\end{corollary}
The weak convergence results given by~(\ref{cv1}) and~(\ref{cv2}), with
respect to the CV ensemble prediction risks, are not surprising. However,
the CV ensemble prediction risks differ from the model prediction
risks, as shown in~(\ref{cvi}) and~(\ref{cvii}), with the class-specific
discrepancies in particular exhibiting a slower decay rate. Consequently, the
CV estimator
$\xbar{\psi}_{d,\mathrm{cv}}$ converges to $\psi_d(\hb)$ at a rate of order
$n^{-1/3}$, slower than the empirical risk $\h{\psi}_d(\hb)$ of order
$n^{-1/2}$ as given by Theorem~\ref{thm3}. This result explains the
observation in Example~\ref{expl1} that $\xbar{\psi}_{d,\mathrm{cv}}$ tracked
$\psi_d(\hb)$ less accurately. For overall risk, the discrepancy is of order
$n^{-2/3}$ and so the CV overall risk
$\xbar{\psi}_{\mathrm{cv}}$ is first-order asymptotically equivalent to the
empirical risk $\h{\psi}(\hb)$. Nevertheless, the CV overall risk exhibits
a positive second-order bias, which decreases as $K$ increases, and tends to
be conservative in contrast to the empirical risk.

\subsection{Proposed $K$-fold CAP}

The theoretical results on the empirical risk and $K$-fold CV
motivate a novel two-step model risk prediction method. The cross-audit step
utilizes the same resampling scheme as $K$-fold CV to estimate the
over-optimism in subsamples. Subsequently, the projection step accounts for
the sample-size reduction to correct bias in the empirical risk.

{\em The Cross-Audit Step.} Write $\h{\psi}_d^{(-k)}({\bf b})$ and
$\h{\psi}^{(-k)}({\bf b})$ as the
counterparts of $\h{\psi}_d({\bf b})$ and $\h{\psi}({\bf b})$, respectively,
using the data except the $k$-th fold, for $k=1,\ldots,K$ and $d=0,1$. Obtain
$K$-fold empirical risks:
\begin{equation}
\xbar{\psi}_{d,\mathrm{emp}}=K^{-1}\sum_{i=1}^K\h{\psi}_d^{(-k)}(\hb^{(-k)}),\quad d=0,1,
\qquad \xbar{\psi}_{\mathrm{emp}}=\omega\xbar{\psi}_{0,\mathrm{emp}}
+(1-\omega)\xbar{\psi}_{1,\mathrm{emp}}.
\label{caprisk}
\end{equation}
Their deviations from the $K$-fold CV counterparts, $\xbar{\psi}_{d,\mathrm{cv}}$
and $\xbar{\psi}_{\mathrm{cv}}$, provide bias estimates of empirical risks in
the subsamples. As a
consequence of ERM, $\xbar{\psi}_{\mathrm{emp}}$
and $\xbar{\psi}_{\mathrm{cv}}$ satisfy an inequality relationship.
\begin{prop}\label{prop1}
  Consider $\xbar{\psi}_{\mathrm{cv}}$ and $\xbar{\psi}_{\mathrm{emp}}$ defined
  in~(\ref{cvrisk}) and~(\ref{caprisk}), respectively. Then,
  \[ \xbar{\psi}_{\mathrm{emp}} \leq \xbar{\psi}_{\mathrm{cv}} + (K-1)\epi_n, \]
  where $\epi_n$ is the error tolerance in the definition~(\ref{est}) of the
  estimated classifier.
\end{prop}
Thus, if $\epi_n$ is set to 0,
the bias estimate of $\xbar{\psi}_{\mathrm{emp}}$ is
nonpositive.

{\em The Projection Step.}
As the subsamples are $(K-1)/K$ of the full sample in size, the bias in
$\xbar{\psi}_{d,\mathrm{emp}}$ is expected to be larger than that in
$\h{\psi}_d(\hb)$. Shrink the bias estimate of $\xbar{\psi}_{d,\mathrm{emp}}$
by a factor of $\{(K-1)/K\}^{2/3}$, to be consistent with the asymptotic
bias of order $n^{-2/3}$ given in Theorem~\ref{thm3}:
\begin{eqnarray}
  \h{\psi}_{d,\mathrm{cap}} &=&
  Q^{-1}\left[Q\{\h{\psi}_d(\hb)\}-\{(K-1)/K\}^{2/3}
     \left\{Q(\xbar{\psi}_{d,\mathrm{emp}})-Q(\xbar{\psi}_{d,\mathrm{cv}})\right\}\right],\label{proj}\\
  \h{\psi}_{\mathrm{cap}} &=& \omega\h{\psi}_{0,\mathrm{cap}}
     +(1-\omega)\h{\psi}_{1,\mathrm{cap}},
\end{eqnarray}
where $Q(\cdot)$ is a given monotone function.
The choice of $Q(\cdot)$ is not
essential asymptotically and could simply be, for example,
the identity function. In our numerical studies, $Q(\cdot)$ was taken to be the
standard normal quantile function for range preservation.

\begin{corollary}\label{coro3}
  Suppose that $K$ is fixed, and monotone function $Q(\cdot)$ is differentiable
  at $\psi_d(\bt)$, $d=0,1$, with non-zero derivatives.
  Under Conditions~\ref{con1}--\ref{con5},
  $n^{2/3}[\{\h{\psi}_{d,\mathrm{cap}}-\psi_d(\hb)\} -
    \{\h{\psi}_d(\bt)-\psi_d(\bt)\}]$,
  $d=0,1$,
  and
  $n^{2/3}[\{\h{\psi}_{\mathrm{cap}}-\psi(\hb)\} -
    \{\h{\psi}(\bt)-\psi(\bt)\}]$
  converge weakly to non-degenerate and mean-zero distributions.
\end{corollary}

For overall risk prediction, the CAP estimator $\h{\psi}_{\mathrm{cap}}$ is
first-order asymptotically equivalent to both the empirical estimator
$\h{\psi}(\hb)$ and the CV estimator $\xbar{\psi}_{\mathrm{cv}}$. However,
the CAP estimator is second-order asymptotically unbiased, whereas the
other two are not. Meanwhile, for class-specific risk prediction, the CAP
estimator $\h{\psi}_{d,\mathrm{cap}}$ and the empirical estimator
$\h{\psi}_d(\hb)$ are first-order asymptotically equivalent, both converging
faster than the CV estimator $\xbar{\psi}_{d,\mathrm{cv}}$. The class-specific
CAP estimator remains second-order asymptotically unbiased.

\begin{remark}\label{rmk1}\rm
  Model prediction risk is random because it depends on the training data, and
  its expectation is the {\em algorithm prediction risk}. Between the two
  estimands, existing estimators in the literature are often found to track
  the latter more closely despite that the former is typically of greater
  interest. In particular, CV estimators behave
  in this way in a number of settings; see \cite{zhang}, \cite{efron97},
  \cite{hastie}, and \cite{bates}. In contrast, the class-specific CAP
  risk estimator tracks the model prediction risk more closely because
  $\h{\psi}_{d,\mathrm{cap}}-\psi_d(\hb)=O_p(n^{-1/2})$ is dominated
  by the $n^{-1/3}$-order variation of the model prediction risk $\psi_d(\hb)$.
  See Supplementary Appendix~B for a numerical verification using
  Example~\ref{expl1}.
\end{remark}

\begin{remark}\label{rmk2}\rm
  As an extension of $K$-fold CV, repeated $K$-fold CV has been suggested to
  stabilize estimation particularly for small samples \citep{burman}. The
  same strategy applies to $K$-fold CAP. With the number of repetitions fixed,
  the asymptotic results in Corollaries~\ref{coro2}--\ref{coro3} extends
  in a straightforward manner to repeated $K$-fold CV and CAP. In particular,
  repeated $K$-fold CAP preserves the same first-order asymptotics and
  second-order asymptotic unbiasedness.
\end{remark}

\begin{remark}\label{rmk3}\rm
  Unlike $K$-fold CV, $K$-fold CAP is largely insensitive to the choice of $K$.
  Corollary~\ref{coro3} shows that
  neither the first-order asymptotics nor the second-order asymptotic
  unbiasedness is affected. Smaller values of $K$ offer computational
  advantages.
  Our numerical experience suggests that repeated 2-fold CAP with 16
  repetitions performs well, which was the CAP method adopted in the simulation
  studies of Example~\ref{expl1}.
\end{remark}

\begin{remark}\label{remarka}\rm
  For computational convenience, classifiers are often constructed in practice
  using logistic regression~(LR) for feature combination, followed by ERM for
  threshold estimation. The estimated combination coefficients typically
  converge at the faster parametric rate, even under model misspecification.
  Consequently, the resulting combination can be treated as a fixed
  single feature in model risk prediction of this LR-ERM classifier.
\end{remark}

\subsection{Inference with model risk bound}

Following Corollary~\ref{coro3},
\[
\left(\begin{array}{c}\h{\psi}_{0,\mathrm{cap}}\\ \h{\psi}_{1,\mathrm{cap}}
\end{array}\right)\sim
\mbox{AN}\left\{\left(\begin{array}{c}\psi_0(\hb)\\ \psi_1(\hb)
\end{array}\right),
\left(\begin{array}{cc}\psi_0(\hb)\{1-\psi_0(\hb)\}/n_0 & 0\\
    0 & \psi_1(\hb)\{1-\psi_1(\hb)\}/n_1\end{array}\right)\right\},
\]
where $\mbox{AN}$ denotes asymptotically normal.
Marginally, the inference of class-specific risk $\psi_d(\hb)$ is analogous
to that of a binomial proportion. Among confidence intervals for a binomial
proportion, the score interval \citep{wilson} is range-preserving and
generally preferred \citep[e.g.,][]{agresti}. We mimic this approach to
derive a upper bound for $\psi_d(\hb)$, with nominal probability
$\delta_d$, as the larger solution $q$ to
\[
\frac{n_d(\h{\psi}_{d,\mathrm{cap}}-q)^2}{q(1-q)}=z_{\delta_d}^2,
\]
where $z_{\delta_d}$ is the $\delta_d$-quantile of the standard normal
distribution. Because of the asymptotic independence, jointly the two
class-specific risk bounds then have nominal probability $\delta_0\delta_1$.
For overall risk $\psi(\hb)$, we propose a upper bound, with nominal
probability $\delta$, as the larger solution $q$ to:
\[
\min_{q_0,q_1:\omega q_0+(1-\omega)q_1=q}
\frac{n_0(\h{\psi}_{0,\mathrm{cap}}-q_0)^2}{q_0(1-q_0)}+
\frac{n_1(\h{\psi}_{1,\mathrm{cap}}-q_1)^2}{q_1(1-q_1)}
=z_{\delta}^2,
\]
on the basis of quadratic inference function \citep[e.g.,][]{lindsay}.

The preceding risk inference is first-order asymptotically justified. Although
this justification remains valid when the empirical risks are used instead in the
procedure, performance may deteriorate due to their substantial bias. On the
other hand, it is unclear how to construct a suitable inference procedure
based on $K$-fold CV, even for the overall risk, despite the fact that the
$K$-fold CV overall risk point estimator is first-order asymptotically
equivalent to its $K$-fold CAP counterpart.

\section{Numerical studies}

Simulations were conducted to evaluate linear ERM classification and model
risk prediction methods for both ERM and LR-ERM classifiers. For illustration,
we also present an application to breast cancer detection. The
algorithmic implementation of linear ERM classification is detailed in
Supplementary Appendix~A.

\subsection{Simulations}

We considered the same set-ups as in \cite{huang22}. In all of them,
features in class~0 were independent and identically distributed as
standard normal, whereas those in class~1 followed distributions that varied
across set-ups. Either three or six features were considered.
For the former, all the features were informative as
their class~1 distributions differed from standard normal. In the latter
case, three additional independent noninformative features were included,
each in class~1 following the standard normal distribution. Four class~1
distributions for the three informative features were formulated, mimicking
cancer detection applications:
\begin{list}{}{\itemsep 0ex\parsep 0ex\topsep 0ex}
\item[{\it Scenario~A:}] independent and identically distributed as normal
  with mean 0.9 and variance~1;
\item[{\it Scenario~B:}] independent and normally distributed with the same
  mean of 0.8 but different variances of 0.5, 1, and 2;
\item[{\it Scenario~C:}] normally distributed with the same mean of 1,
  different variances of 0.5, 1, and 2, and pairwise correlation coefficient
  of 0.5;
\item[{\it Scenario~D:}] mixture of two distributions each with the three
  features being independent and normally distributed, one with probability
  2/3 having means of 1.7, 1.7, and 0 and variances of 0.5, 2, and 1,
  and the other with probability 1/3 having means
  of 0, 0, and 1.7 and the same variance of 1.
\end{list}
In the first three scenarios, class~1 features differed in their dependence
structure and variability. In contrast, Scenario~D emulated cancer
heterogeneity, with two subtypes associated with elevations in the first
two features and the last feature separately. The overall risk used
$\omega=0.25$, and the optimal classifiers yielded class-specific risks of
(0.47, 0.069), (0.54, 0.052), (0.57, 0.045), and (0.48, 0.072) for classes~0
and~1 under Scenarios~A--D, respectively. The class sizes were set equal,
$n_0 = n_1$, ranging from 100 to 500. Results were based on 1000 simulated
datasets for each set-up.

\begin{sidewaystable}[htbp]
  \caption{\label{tab1}Simulation results on overall $|$ class~0 $|$ class~1
    risk prediction of the ERM classifier with three features}
\centering
\begin{tabular}{
    c@{\hspace*{8pt}}
    r@{$|$}r@{$|$}r@{\hspace*{6pt}}r@{$|$}r@{$|$}r@{\hspace*{8pt}}
    r@{$|$}r@{$|$}r@{\hspace*{6pt}}r@{$|$}r@{$|$}r@{\hspace*{8pt}}
    r@{$|$}r@{$|$}r@{\hspace*{6pt}}r@{$|$}r@{$|$}r@{\hspace*{8pt}}
    r@{$|$}r@{$|$}r@{\hspace*{6pt}}r@{$|$}r@{$|$}r@{\hspace*{6pt}}r@{$|$}r@{$|$}r}\\\hline\hline\\[-10pt]
  $n_0\phantom{=}$ & \multicolumn{6}{c}{empirical} & \multicolumn{6}{c}{bootstrap} & \multicolumn{6}{c}{10-fold CV}
  & \multicolumn{9}{c}{repeated 2-fold CAP}\\
  \cmidrule(lr){2-7}\cmidrule(lr){8-13}\cmidrule(lr){14-19}\cmidrule(lr){20-28}\\[-10pt]
  $\phantom{,}=n_1$ &
  \multicolumn{3}{c}{B} & \multicolumn{3}{c}{D} & \multicolumn{3}{c}{B} &
  \multicolumn{3}{c}{D} & \multicolumn{3}{c}{B} & \multicolumn{3}{c}{D} &
  \multicolumn{3}{c}{B} & \multicolumn{3}{c}{D} & \multicolumn{3}{c}{C}
  \\\hline\\[-10pt]
  \multicolumn{28}{c}{Scenario A}\\
100 &$-$38 &$-$42 &$-$36 &38 &51 &38 &$-$14 &$-$16 &$-$13 &22 &40 &24 & 1 &$-$4 & 3 &23 &51 &30 &$-$2 &$-$3 &$-$2 &20 &39 &24 &94.4 &94.9 &94.0\\
200 &$-$25 &$-$25 &$-$25 &26 &34 &26 &$-$8 &$-$8 &$-$8 &15 &28 &17 & 0 &$-$2 & 1 &16 &38 &21 &$-$1 & 0 &$-$1 &13 &27 &16 &93.9 &95.9 &93.1\\
300 &$-$20 &$-$20 &$-$20 &20 &28 &21 &$-$6 &$-$7 &$-$6 &12 &24 &13 &0 &0 &0 &13 &33 &17 &$-$1 &$-$1 &$-$1 &11 &22 &12 &94.4 &94.6 &93.9\\
400 &$-$17 &$-$18 &$-$16 &17 &24 &17 &$-$5 &$-$5 &$-$5 &10 &20 &11 & 0 &$-$2 & 1 &10 &29 &14 &$-$1 &$-$1 &$-$1 &10 &20 &11 &94.6 &94.4 &95.0\\
500 &$-$14 &$-$14 &$-$14 &15 &21 &15 &$-$5 &$-$5 &$-$4 & 9 &18 &10 &1 &0 &1 & 9 &26 &13 & 0 & 0 &$-$1 & 8 &18 & 9 &94.3 &95.6 &95.0\\
  \multicolumn{28}{c}{Scenario B}\\
100 &$-$36 &$-$37 &$-$35 &36 &48 &36 &$-$13 &$-$15 &$-$13 &21 &41 &22 & 0 &$-$5 & 2 &21 &51 &27 &$-$2 & 0 &$-$2 &20 &39 &23 &92.0 &95.5 &93.3\\
200 &$-$24 &$-$23 &$-$25 &25 &33 &26 &$-$8 &$-$7 &$-$8 &14 &28 &15 & 1 &$-$1 & 2 &16 &37 &19 &$-$2 & 1 &$-$2 &14 &28 &16 &92.6 &95.0 &91.7\\
300 &$-$19 &$-$19 &$-$19 &19 &27 &20 &$-$6 &$-$6 &$-$6 &11 &23 &12 &0 &0 &0 &12 &31 &15 &$-$1 &$-$1 &$-$1 &11 &23 &12 &94.6 &94.8 &93.9\\
400 &$-$16 &$-$16 &$-$15 &16 &23 &16 &$-$5 &$-$5 &$-$5 &10 &20 &10 & 1 &$-$2 & 1 &10 &28 &13 &$-$1 &$-$2 &$-$1 & 9 &19 &10 &94.4 &94.2 &94.5\\
500 &$-$14 &$-$13 &$-$14 &14 &21 &14 &$-$4 &$-$4 &$-$4 & 8 &18 & 9 &1 &0 &1 & 9 &25 &11 &$-$1 &$-$1 &$-$1 & 8 &18 & 9 &93.5 &94.3 &94.7\\
  \multicolumn{28}{c}{Scenario C}\\
100 &$-$37 &$-$44 &$-$34 &37 &53 &36 &$-$12 &$-$14 &$-$12 &22 &40 &24 & 0 &$-$1 & 0 &24 &52 &30 &$-$3 &$-$3 &$-$2 &20 &40 &24 &94.9 &94.9 &93.7\\
200 &$-$24 &$-$27 &$-$23 &24 &35 &24 &$-$6 &$-$6 &$-$6 &14 &28 &15 & 1 &$-$3 & 2 &16 &41 &21 &$-$2 &$-$1 &$-$2 &14 &29 &16 &93.5 &94.2 &94.1\\
300 &$-$18 &$-$19 &$-$17 &18 &28 &18 &$-$4 &$-$4 &$-$4 &11 &23 &12 & 1 &$-$1 & 2 &12 &35 &17 &$-$1 & 0 &$-$2 &11 &23 &12 &93.9 &95.0 &94.4\\
400 &$-$14 &$-$16 &$-$14 &15 &23 &15 &$-$4 &$-$3 &$-$4 &10 &20 &11 & 1 &$-$3 & 2 &11 &30 &15 &$-$1 & 0 &$-$1 & 9 &20 &10 &94.1 &95.0 &94.4\\
500 &$-$13 &$-$14 &$-$12 &14 &21 &14 &$-$3 &$-$3 &$-$3 & 8 &18 & 9 & 1 &$-$2 & 2 & 9 &29 &13 &$-$1 & 0 &$-$1 & 9 &18 &10 &94.9 &94.2 &92.6\\
  \multicolumn{28}{c}{Scenario D}\\
100 &$-$40 &$-$33 &$-$42 &40 &46 &43 &$-$13 &$-$10 &$-$15 &23 &40 &26 &2 &0 &3 &24 &54 &31 &$-$2 & 1 &$-$3 &22 &38 &27 &93.7 &95.6 &92.2\\
200 &$-$26 &$-$22 &$-$27 &26 &31 &28 &$-$9 &$-$7 &$-$9 &16 &28 &18 & 1 &$-$3 & 2 &16 &41 &21 &$-$1 & 0 &$-$1 &14 &27 &17 &94.3 &95.4 &94.7\\
300 &$-$20 &$-$17 &$-$21 &21 &27 &22 &$-$6 &$-$5 &$-$7 &12 &23 &14 & 0 &$-$2 & 1 &13 &32 &17 &$-$1 &$-$1 &$-$1 &12 &23 &13 &94.0 &94.4 &94.5\\
400 &$-$17 &$-$14 &$-$18 &18 &23 &19 &$-$5 &$-$5 &$-$5 &10 &20 &12 &1 &0 &1 &11 &28 &15 &$-$1 & 0 &$-$1 &10 &20 &11 &93.9 &94.5 &94.3\\
500 &$-$15 &$-$12 &$-$16 &15 &21 &16 &$-$4 &$-$4 &$-$5 & 9 &18 &10 &1 &0 &1 &10 &28 &14 &$-$1 & 0 &$-$1 & 9 &18 &10 &94.7 &95.4 &94.9\\
\hline\\[-10pt]
\multicolumn{28}{l}{\footnotesize
  B: bias ($\times1000$), D: mean absolute
  deviation ($\times1000$), C: coverage probability of 95\% risk bound (\%).}
\end{tabular}
\end{sidewaystable}

\begin{sidewaystable}[htbp]
  \caption{\label{tab2}Simulation results on overall $|$ class~0 $|$ class~1
    risk prediction of the ERM classifier with six features}
\centering
\begin{tabular}{
    c@{\hspace*{8pt}}
    r@{$|$}r@{$|$}r@{\hspace*{6pt}}r@{$|$}r@{$|$}r@{\hspace*{8pt}}
    r@{$|$}r@{$|$}r@{\hspace*{6pt}}r@{$|$}r@{$|$}r@{\hspace*{8pt}}
    r@{$|$}r@{$|$}r@{\hspace*{6pt}}r@{$|$}r@{$|$}r@{\hspace*{8pt}}
    r@{$|$}r@{$|$}r@{\hspace*{6pt}}r@{$|$}r@{$|$}r@{\hspace*{6pt}}r@{$|$}r@{$|$}r}\\\hline\hline\\[-10pt]
  $n_0\phantom{=}$ & \multicolumn{6}{c}{empirical} & \multicolumn{6}{c}{bootstrap} & \multicolumn{6}{c}{10-fold CV}
  & \multicolumn{9}{c}{repeated 2-fold CAP}\\
  \cmidrule(lr){2-7}\cmidrule(lr){8-13}\cmidrule(lr){14-19}\cmidrule(lr){20-28}\\[-10pt]
  $\phantom{,}=n_1$ &
  \multicolumn{3}{c}{B} & \multicolumn{3}{c}{D} & \multicolumn{3}{c}{B} &
  \multicolumn{3}{c}{D} & \multicolumn{3}{c}{B} & \multicolumn{3}{c}{D} &
  \multicolumn{3}{c}{B} & \multicolumn{3}{c}{D} & \multicolumn{3}{c}{C}
  \\\hline\\[-10pt]
  \multicolumn{28}{c}{Scenario A}\\
100 &$-$59 &$-$63 &$-$57 &59 &67 &58 &$-$23 &$-$20 &$-$24 &28 &43 &31 & 0 &$-$8 & 3 &25 &52 &31 &$-$5 &$-$1 &$-$6 &24 &41 &29 &89.2 &93.5 &88.2\\
200 &$-$39 &$-$41 &$-$38 &39 &45 &38 &$-$13 &$-$13 &$-$13 &17 &30 &19 & 0 &$-$2 & 1 &17 &37 &21 &$-$3 &$-$1 &$-$3 &15 &29 &17 &91.9 &93.5 &92.3\\
300 &$-$30 &$-$31 &$-$30 &30 &34 &30 &$-$10 &$-$10 & $-$9 &13 &24 &15 &0 &1 &0 &12 &33 &17 &$-$2 & 0 &$-$3 &12 &23 &14 &92.7 &94.2 &92.6\\
400 &$-$25 &$-$26 &$-$24 &25 &30 &25 &$-$8 &$-$7 &$-$8 &12 &21 &13 & 0 &$-$2 & 1 &11 &27 &15 &$-$1 &$-$1 &$-$1 &10 &20 &12 &92.4 &94.4 &93.3\\
500 &$-$20 &$-$21 &$-$20 &20 &24 &21 &$-$7 &$-$6 &$-$7 &10 &18 &11 &1 &0 &1 & 9 &26 &13 &0 &1 &0 & 8 &17 & 9 &95.8 &95.7 &95.0\\
  \multicolumn{28}{c}{Scenario B}\\
100 &$-$57 &$-$59 &$-$57 &58 &65 &57 &$-$22 &$-$19 &$-$22 &27 &44 &29 & 2 &$-$5 & 4 &24 &50 &31 &$-$6 & 0 &$-$8 &24 &41 &29 &87.2 &93.5 &86.5\\
200 &$-$37 &$-$38 &$-$37 &37 &43 &37 &$-$13 &$-$11 &$-$13 &17 &31 &18 &1 &1 &1 &16 &37 &20 &$-$3 &$-$1 &$-$4 &15 &28 &17 &90.8 &94.2 &91.5\\
300 &$-$29 &$-$28 &$-$30 &29 &32 &30 &$-$9 &$-$9 &$-$9 &13 &24 &14 &0 &0 &0 &12 &32 &16 &$-$3 & 1 &$-$4 &11 &24 &13 &92.9 &94.4 &92.4\\
400 &$-$24 &$-$24 &$-$23 &24 &28 &24 &$-$7 &$-$6 &$-$8 &11 &21 &12 & 0 &$-$1 & 1 &11 &28 &14 &$-$1 & 0 &$-$2 & 9 &20 &11 &93.5 &94.9 &92.9\\
500 &$-$19 &$-$19 &$-$20 &20 &23 &20 &$-$7 &$-$5 &$-$7 &10 &18 &10 & 1 &$-$1 & 1 & 9 &24 &12 &0 &1 &0 & 8 &17 & 9 &94.3 &95.9 &95.1\\
  \multicolumn{28}{c}{Scenario C}\\
100 &$-$58 &$-$66 &$-$56 &58 &70 &56 &$-$21 &$-$19 &$-$21 &26 &43 &28 & 2 &$-$1 & 3 &25 &55 &33 &$-$6 &$-$2 &$-$8 &24 &40 &29 &90.8 &94.8 &89.5\\
200 &$-$37 &$-$40 &$-$36 &37 &44 &36 &$-$12 &$-$12 &$-$11 &17 &30 &18 & 1 &$-$2 & 2 &17 &40 &23 &$-$3 & 1 &$-$4 &16 &28 &18 &90.9 &94.6 &90.7\\
300 &$-$27 &$-$30 &$-$27 &28 &34 &27 & $-$9 & $-$9 &$-$10 &13 &24 &14 & 1 &$-$1 & 2 &12 &33 &17 &$-$2 & 1 &$-$3 &12 &24 &13 &93.8 &94.1 &92.5\\
400 &$-$23 &$-$24 &$-$22 &23 &28 &23 &$-$7 &$-$7 &$-$7 &11 &21 &12 & 0 &$-$2 & 0 &11 &30 &16 &$-$1 & 2 &$-$2 & 9 &20 &11 &94.9 &95.0 &93.5\\
500 &$-$20 &$-$21 &$-$19 &20 &25 &20 &$-$5 &$-$5 &$-$5 & 9 &18 &10 & 0 &$-$1 & 1 &10 &27 &14 &$-$1 & 1 &$-$2 & 9 &18 &10 &94.5 &95.0 &93.5\\
  \multicolumn{28}{c}{Scenario D}\\
100 &$-$59 &$-$56 &$-$61 &59 &62 &61 &$-$24 &$-$20 &$-$25 &29 &44 &32 & 2 &$-$1 & 2 &25 &53 &32 &$-$4 & 2 &$-$7 &24 &39 &29 &90.4 &94.8 &88.6\\
200 &$-$38 &$-$34 &$-$40 &39 &40 &40 &$-$13 & $-$9 &$-$14 &18 &28 &20 & 0 &$-$3 & 1 &17 &38 &23 &$-$2 & 3 &$-$4 &16 &29 &18 &93.3 &96.1 &91.9\\
300 &$-$31 &$-$29 &$-$32 &31 &32 &32 & $-$9 & $-$9 &$-$10 &14 &24 &15 & 0 &$-$3 & 1 &13 &34 &18 &$-$2 & 0 &$-$3 &12 &22 &14 &93.3 &95.1 &94.4\\
400 &$-$25 &$-$24 &$-$26 &25 &28 &26 &$-$8 &$-$8 &$-$8 &12 &21 &14 &0 &0 &0 &12 &29 &16 &$-$1 & 0 &$-$2 &10 &20 &12 &93.1 &94.9 &92.8\\
500 &$-$21 &$-$20 &$-$22 &21 &24 &22 &$-$7 &$-$5 &$-$7 &10 &18 &11 &1 &0 &1 &10 &26 &14 &0 &0 &0 & 9 &18 &11 &94.6 &94.5 &95.4\\
\hline\\[-10pt]
\multicolumn{28}{l}{\footnotesize
  B: bias ($\times1000$), D: mean absolute
  deviation ($\times1000$), C: coverage probability of 95\% risk bound (\%).}
\end{tabular}
\end{sidewaystable}

The main focus was on linear ERM classification. Our algorithm proved both
effective and efficient, and the estimated classifier behaved in accordance
with the asymptotic theory; see Supplementary Appendix~C.
For the model risk prediction, Tables~\ref{tab1} and~\ref{tab2} report results
for repeated 2-fold CAP with 16 repetitions along with empirical risk
estimation, the bootstrap method of \citet{harrell} with 100 resamples, and
10-fold CV, for settings with three and six features, respectively. As expected,
the empirical estimator exhibited substantial over-optimism for both overall and
class-specific risks. The bootstrap estimator reduced this bias but
remained overly optimistic. In contrast, 10-fold CV exhibited negligible bias.
While the mean absolute deviation of 10-fold CV for overall
risk was smaller than that of
the empirical estimator and
comparable to that of the bootstrap estimator, this pattern did not extend
to class-specific risks, particularly with larger sample sizes, mirroring
the phenomenon observed in Example~\ref{expl1}. By comparison,
repeated 2-fold CAP outperformed the alternatives for both overall and
class-specific risk in terms of bias and mean absolute deviation. Moreover,
its risk bound had coverage close to the nominal level, especially as the
sample size increased. Additional results on repeated 2-fold CV, repeated
10-fold CV, and repeated 10-fold CAP further confirmed the superior
performance of $K$-fold CAP in general and its relative insensitivity to the
choice of $K$; see Supplementary Appendix~C. Simulations for the model
risk prediction of LR-ERM
classification, which is more commonly used in practice (Remark~\ref{remarka}),
showed similar performance patterns; see Supplementary
Appendix~D.

\subsection{Application to breast cancer detection}

For illustration, we analyzed a clinical study for breast cancer detection
using demographic variables and routine blood measurements \citep{patricio}.
The Breast Cancer Coimbra dataset \citep{patriciodata} consists of 64
women with breast cancer and 52 healthy volunteers. The features
considered were glucose, resistin, age, and BMI; all were
log-transformed. To prioritize accurate classification of cancer cases, we
set $\omega = 0.25$.

\begin{table}[b!]
  \caption{\label{tab3}ERM and LR-ERM classifications for breast cancer
  detection}
\centering
\begin{tabular}{lrrclr@{$|$}r@{$|$}r r@{$|$}r@{$|$}r}\\\hline\hline\\[-10pt]
          & \multicolumn{1}{c}{ERM} & \multicolumn{1}{c}{LR-ERM} &&&  \multicolumn{3}{c}{ERM} & \multicolumn{3}{c}{LR-ERM}\\\hline\\[-10pt]
\multicolumn{3}{c}{estimated classifiers} && \multicolumn{7}{c}{model risk prediction}\\\\[-15pt]
  &     &        &&& overall & HL & BC & overall & HL & BC\\
threshold     &  0.436 &  0.661 && emp     & 0.148 & 0.500 & 0.031 & 0.173 & 0.365 & 0.109\\
log(glucose)  &  0.273 &  0.207 && boot    & 0.192 & 0.580 & 0.063 & 0.203 & 0.418 & 0.131\\
log(resistin) &  0.052 &  0.031 && CV      & 0.233 & 0.558 & 0.125 & 0.214 & 0.481 & 0.125\\
log(age)      & $-0.145$ & $-0.012$ && CAP     & 0.217 & 0.612 & 0.085 & 0.223 & 0.429 & 0.154\\
log(BMI)      & $-0.093$ & $-0.088$ && RB & 0.278 & 0.715 & 0.161 & 0.294 & 0.542 & 0.242\\
\hline\\[-10pt]
\multicolumn{11}{l}{\footnotesize
  emp: empirical estimation, boot: bootstrap of Harrell et~al.\ with 100 re-samples, CV: 10-fold CV,}\\
\multicolumn{11}{l}{\footnotesize
  CAP: repeated 2-fold CAP with 16 repetitions, RB: 95\% risk bound.}\\
\multicolumn{10}{l}{\footnotesize
  HL: healthy volunteer, BC: breast cancer.}
\end{tabular}
\end{table}

Table~\ref{tab3} reports the ERM and LR-ERM classifiers together with their
estimated model prediction risks. The two classifiers differed considerably.
Although LR-ERM classifier is often used for computational convenience, it
can be less predictive than ERM classifier under model misspecification.
Indeed, the
CAP risk estimates, along with the 95\% risk bounds, suggested improved
performance of the ERM classifier. The four point prediction risk estimates
were broadly consistent with the patterns observed in the simulation studies.

\section{Discussion}

In this article, we have developed $K$-fold CAP, a novel and theoretically
justified method for model risk prediction. Like classical analytic correction
methods, it aims to correct the over-optimism of empirical risk, but does
so via a $K$-fold CV–style resampling scheme informed by asymptotic theory.
For concreteness, the exposition has focused on linear ERM classification.
Nevertheless, $K$-fold CAP applies generally to prediction algorithms
involving cube-root asymptotics. The generalization can be made even more
broadly, by matching the shrinking factor in the projection step with the
order of the leading asymptotic bias of the empirical risk. For most problems
amenable to classical analytic correction methods, including prediction
likelihood targeted by AIC, the bias is of order $n^{-1}$ leading to a
shrinking factor of $(K-1)/K$. However, the asymptotic bias for many modern
and complex prediction algorithms remains poorly understood, which merits
further investigation.

A fully automatic procedure that does not require asymptotic analysis
would be desirable. However, the asymptotic bias, which typically
arises in second-order asymptotics, may not have its order reliably estimated
from the data. Nevertheless, $K$-fold CAP can adopt a {\em working} shrinking
factor. Although second-order
asymptotic unbiasedness is no
longer guaranteed, the procedure still possesses theoretical advantages over
$K$-fold CV and could achieve a substantial bias reduction. Suppose that the
over-optimism decays no faster than order $n^{-1}$, as faster decay would
make it less consequential. If we adopt a working shrinking factor
of $\{(K-1)/K\}^{1/2}$, then $10$-fold CAP incurs a second-order asymptotic
bias within $1-(9/10)^{\pm1/2}\approx\pm5\%$ of that of the empirical risk.

While model prediction risk is often targeted in practice, other notions
of prediction risk may also be of interest, particularly algorithm prediction
risk; see Remark~\ref{rmk1}. Defined as the expectation of model
prediction risk, it reflects the performance of the learning
algorithm rather than a specific learned model; see \citet{dietterich}
for a detailed discussion of prediction estimands and their practical
implications. $K$-fold CV has been used to estimate algorithm
prediction risk, with its variance studied for uncertainty quantification
\citep{nadeau,bengio,markatou}. Our results suggest that
$K$-fold CAP may also offer advantages for this purpose. Further investigation,
along with the development of the associated inferential procedures, is
an important direction for future work.

\section*{Appendix: Proofs}

\subsubsection*{Proof of Theorem~\ref{thm1}}
Under Condition~\ref{con1}, almost surely
\[
\sup_{{\bf b}}|\h{\psi}_d({\bf b})-\psi_d({\bf b})|=o(1),\qquad d=0,1,
\]
since $\{I({\bf b}^\T{\bf X}_0^\circ>0):{\bf b}\in\mathbb{R}^{m+1}\}$ and
$\{I({\bf b}^\T{\bf X}_1^\circ\leq0):{\bf b}\in\mathbb{R}^{m+1}\}$
are Donsker classes \citep[e.g.,][lemma~9.12]{kosorok} and thus also
Glivenko--Cantelli classes. Subsequently,
\[
\sup_{{\bf b}}|\h{\psi}({\bf b})-\psi({\bf b})|=o(1),
\]
almost surely. Therefore, $\h{\psi}(\hb)-\psi(\hb)=o(1)$ and
$\h{\psi}(\bt)-\psi(\bt)=o(1)$, almost surely. They, together with
$\h{\psi}(\hb)\leq\h{\psi}(\bt)+o(1)$ implied by display~(\ref{est}),
give $\psi(\hb)\leq\psi(\bt)+o(1)$ almost surely. Under Condition~\ref{con2},
the strong consistency of $\hb$ then follows standard $M$-estimation
arguments \citep[e.g.,][proof of theorem~5.7]{vdv}.

\subsubsection*{Proof of Theorem~\ref{thm2}}
The proof follows the general cube-root asymptotics framework of \cite{kim},
with two adaptations for our problem, the case-control design and
minimization formulation.
Write
\[
g({\bf x}^\circ;{\bf b}_{-\ell})
=I({\bf b}_{-\ell}^\T{\bf x}_{-\ell}^\circ+\bt_\ell {\bf x}^\circ_\ell\leq0)
-I(\bt^\T{\bf x}^\circ\leq0).
\]
Note that $\h{\phi}_d({\bf b}_{-\ell})-\h{\phi}_d(\bt_{-\ell})
=(-1)^{d+1}\he g({\bf X}_d^\circ;{\bf b}_{-\ell})$, for $d=0,1$.
The class of functions,
\[
\mathscr{G}_\epi=\{g(\cdot;{\bf b}_{-\ell}):
\|{\bf b}_{-\ell}-\bt_{-\ell}\|_\infty\leq\epi\},
\]
has envelope
\begin{eqnarray*}
  G_\epi({\bf x}^\circ) &=& \sup_{\|{\bf b}_{-\ell}-\bt_{-\ell}\|_\infty\leq\epi}
  |g({\bf x}^\circ;{\bf b}_{-\ell})|\\
  &=& I\left\{\min_{\|{\bf b}_{-\ell}-\bt_{-\ell}\|_\infty\leq\epi}
  (\bt_{-\ell}-{\bf b}_{-\ell})^\T{\bf x}^\circ_{-\ell}
  <\bt^\T{\bf x}^\circ
  \leq
  \max_{\|{\bf b}_{-\ell}-\bt_{-\ell}\|_\infty\leq\epi}
  (\bt_{-\ell}-{\bf b}_{-\ell})^\T{\bf x}^\circ_{-\ell}\right\}.
\end{eqnarray*}
Since the subgraphs of functions in $\mathscr{G}_\infty$ form a VC class with
bounded $G_\epi({\bf x}^\circ)$, $\mathscr{G}_\epi$ is uniformly manageable
\citep[section~3]{kim}.

We now check the remaining regularity conditions of the main theorem in
\cite{kim}, except for trivial ones due to bounded
$g({\bf x}^\circ;{\bf b}_{-\ell})$ and $G_\epi({\bf x}^\circ)$ in our problem.
Condition~\ref{con4} implies that
$\nabla^2\phi(\bt_{-\ell})$ exists.
Meanwhile,
\begin{eqnarray*}
  V_d({\bf s},{\bf t}) &=& \lim_{\alpha\rightarrow\infty}\alpha
  E\left\{g({\bf X}_d^\circ;\bt_{-\ell}+{\bf s}/\alpha)
  g({\bf X}_d^\circ;\bt_{-\ell}+{\bf t}/\alpha)\right\}\\
  &=& \left\{\begin{array}{ll}\displaystyle
  E\{\min(|{\bf s}^\T{\bf X}_{d,-\ell}^\circ|,
  |{\bf t}^\T{\bf X}_{d,-\ell}^\circ|)
  f_{\bt^\T{\bf X}_d^\circ}(0)
  I({\bf s}^\T{\bf X}_{d,-\ell}^\circ{\bf t}^\T{\bf X}_{d,-\ell}^\circ>0)\}
  & m=1\\[8pt]
  EE\{\min(|{\bf s}^\T{\bf X}_{d,-\ell}^\circ|,
  |{\bf t}^\T{\bf X}_{d,-\ell}^\circ|)
  f_{\bt^\T{\bf X}_d^\circ\mid{\bf X}_{d,-\ell}}(0)\\[8pt]
  \hspace*{1.4in}\times I({\bf s}^\T{\bf X}_{d,-\ell}^\circ{\bf t}^\T{\bf X}_{d,-\ell}^\circ>0)
  \mid{\bf X}_{d,-\ell}\}
  & m\geq2
  \end{array}\right.,
\end{eqnarray*}
exists under Condition~\ref{con3}.
Above, $f_{\bt^\T{\bf X}_d^\circ}$ is the density function of
$\bt^\T{\bf X}_d^\circ$, and $f_{\bt^\T{\bf X}_d^\circ\mid{\bf X}_{d,-\ell}}$ is
the conditional counterpart given ${\bf X}_{d,-\ell}$.

Next, we show $EG_\epi({\bf X}_d^\circ)^2=EG_\epi({\bf X}_d^\circ)=O(\epi)$ as
$\epi\downarrow0$, under Condition~\ref{con3}. This is obvious in the case of
$m=1$. For $m\geq2$,
\begin{eqnarray*}
  EG_\epi({\bf X}_d^\circ) &=& EE\{G_\epi({\bf X}_d^\circ)\mid{\bf X}_{d,-\ell}\}\\
  &=& E\{F_{\bt^\T{\bf X}_d^\circ\mid{\bf X}_{d,-\ell}}(\epi\|{\bf X}_{d,-\ell}^\circ\|_1)
  -F_{\bt^\T{\bf X}_d^\circ\mid{\bf X}_{d,-\ell}}(-\epi\|{\bf X}_{d,-\ell}^\circ\|_1)\}
  \\
  &=& 2\epi E\{f_{\bt^\T{\bf X}_d^\circ\mid{\bf X}_{d,-\ell}}(\epi^*\|{\bf X}_{d,-\ell}^\circ\|_1)\|{\bf X}_{d,-\ell}^\circ\|_1\}\\
  &=& O(\epi),
\end{eqnarray*}
where $F_{\bt^\T{\bf X}_d^\circ\mid{\bf X}_{d,-\ell}}$ is the conditional
distribution function of $\bt^\T{\bf X}_d^\circ$ given
${\bf X}_{d,-\ell}$ and $\epi^*\in[-\epi,\epi]$. Similar arguments can be
used to establish $E|g({\bf X}_d^\circ;{\bf b}_{-\ell})-g({\bf X}_d^\circ;{\bf b}_{-\ell}^*)|=O(\|{\bf b}_{-\ell}-{\bf b}_{-\ell}^*\|_\infty)$ for ${\bf b}_{-\ell}$ and ${\bf b}_{-\ell}^*$ near
$\bt_{-\ell}$.

Then, the assertion on the weak convergence of $\h{\phi}_d(\cdot)$ and
$\h{\phi}(\cdot)$
follows the main theorem of \cite{kim}. Furthermore,
$\nabla^2\phi(\bt_{-\ell})$ is
positive definite under Condition~\ref{con4}. Meanwhile, the
covariance kernel of process $W(\cdot)$, as given by
$V({\bf s},{\bf t})=\omega^2V_0({\bf s},{\bf t})
+(1-\omega)^2V_1({\bf s},{\bf t})$, satisfies
\[
V({\bf s},{\bf s})-2V({\bf s},{\bf t})+V({\bf t},{\bf t})
=V({\bf s}-{\bf t},{\bf s}-{\bf t}).
\]
Then, Gaussian process $Z(\cdot)$ has nondegenerate increments since
$V({\bf s},{\bf s})=\omega^2|{\bf s}^\T\nabla\phi_0(\bt_{-\ell})|
  +(1-\omega)^2|{\bf s}^\T\nabla\phi_1(\bt_{-\ell})|\neq0$ for ${\bf s}\neq{\bf 0}$ under Condition~\ref{con5}. The weak convergence of $\cb_{-\ell}$ subsequently follows.

\subsubsection*{Proof of Theorem~\ref{thm3}}
Following the proof of the argmax continuous mapping theorem in
\citet[theorem~2.7]{kim}, we invoke the almost sure representation theorem
of Dudley for the weak convergence result in~(\ref{wc})
such that the counterparts in a new probability space sharing the same
distributions have almost sure convergence in the sense of
\citet[section~2.2]{kim}:
\[
n^{2/3}\left\{\begin{array}{c}\displaystyle
  \td{\phi}_0(\bt_{-\ell}+n^{-1/3}{\bf s})-\phi_0(\bt_{-\ell}+n^{-1/3}{\bf s})
  -\td{\phi}_0(\bt_{-\ell})+\phi_0(\bt_{-\ell})\\
  \td{\phi}_1(\bt_{-\ell}+n^{-1/3}{\bf s})-\phi_1(\bt_{-\ell}+n^{-1/3}{\bf s})
  -\td{\phi}_1(\bt_{-\ell})+\phi_1(\bt_{-\ell})\\
  \td{\phi}(\bt_{-\ell}+n^{-1/3}{\bf s})-\td{\phi}(\bt_{-\ell})
\end{array}\right\}\rightarrow
\left\{\begin{array}{c}\displaystyle
  \td{W}_0({\bf s})\\
  \td{W}_1({\bf s})\\
  \td{Z}({\bf s})\end{array}\right\},
\]
almost surely. Meanwhile, the processes on the left-hand side are
stochastically equicontinuous \citep[lemma~4.6]{kim}.
Write
$\td{\bt}_{-\ell}$ and $\td{\bf U}$ as the counterparts of $\cb_{-\ell}$ and
$\bf U$, respectively. Similar to \citet[proof of theorem~2.7]{kim}, it can be
established that
\[
\td{\Pr}^*\{\|n^{1/3}(\td{\bt}_{-\ell}-\bt_{-\ell})-\td{\bf U}\|_2>\varepsilon\}\rightarrow0
\]
for each $\varepsilon>0$, where $\td{\Pr}^*$ denotes the outer probability.
Then, it follows that
\[
\td{\Pr}^*\{|n^{2/3}[\{\td{\phi}_d(\td{\bt}_{-\ell})-\phi_d(\td{\bt}_{-\ell})\} - \{\td{\phi}_d(\bt_{-\ell})-\phi_d(\bt_{-\ell})\}]-\td{W}_d(\td{\bf U})|>\varepsilon\}\rightarrow0,\qquad d=0,1,
\]
for each $\varepsilon>0$. Thus,
\begin{eqnarray*}
  n^{2/3}[\{\h{\phi}_d(\cb_{-\ell})-\phi_d(\cb_{-\ell})\}
    - \{\h{\phi}_d(\bt_{-\ell})-\phi_d(\bt_{-\ell})\}]
  &\rightsquigarrow& W_d({\bf U}),\qquad d=0,1,\\
  n^{2/3}[\{\h{\phi}(\cb_{-\ell})-\phi(\cb_{-\ell})\}
    - \{\h{\phi}(\bt_{-\ell})-\phi(\bt_{-\ell})\}]
  &\rightsquigarrow& W({\bf U}).
\end{eqnarray*}
Convergence results in~(\ref{opt1})--(\ref{opt2}) then follow, as
$\h{\phi}_d(\cb_{-\ell})=\h{\psi}_d(\hb)$ and
$\phi_d(\cb_{-\ell})=\psi_d(\hb)$ with probability tending to 1.

Next, we establish that $\bf U$ possesses finite moments of all orders, and
$EW({\bf U})$ exists.
Define a norm $\|{\bf s}\|\equiv({\bf s}^\T{\bf H}{\bf s}/2)^{1/2}$ for
vector $\bf s$. We shall exploit the rescaling property of $W({\bf s})$, that
is, $W(c{\bf s})$ having the same distribution as $c^{1/2}W({\bf s})$ for any
$c>0$. From \citet[lemma~A.2]{shi}, we have
$\Pr\{\inf_{\|{\bf s}\|\leq1}Z({\bf s})\leq -x\}=O\{\exp(-x)\}$
as $x\rightarrow\infty$. It follows
$\Pr\{\inf_{\|{\bf s}\|\leq1}W({\bf s})\leq -x\}\leq c\exp(-x)$
for some constant $c>0$, when $x$ is sufficiently large. Taking such an
$x\geq2$, we have
\begin{eqnarray*}
  \Pr(\|{\bf U}\|>x) &\leq& \Pr\left\{\textstyle\inf_{\|{\bf s}\|>x}Z({\bf s})\leq0\right\}\\
  &\leq& \textstyle\sum_{j=1}^\infty\Pr\left\{\inf_{\|{\bf s}\|\in(x+j-1,x+j]}W({\bf s})+\|{\bf s}\|^2\leq0\right\}\\
  &\leq& \textstyle\sum_{j=1}^\infty\Pr\left[\inf_{\|{\bf s}\|\leq1}W\{(x+j){\bf s}\}\leq -(x+j-1)^2\right]\\
  &=& \textstyle\sum_{j=1}^\infty\Pr\left\{\inf_{\|{\bf s}\|\leq1}W({\bf s})\leq -(x+j-1)^2(x+j)^{-1/2}\right\}\\
  &\leq& \textstyle\sum_{j=1}^\infty\Pr\left\{\inf_{\|{\bf s}\|\leq1}W({\bf s})\leq -x-j+1\right\}\\
  &\leq& \textstyle\sum_{j=1}^\infty c\exp(-x-j+1)\\
  &=& ce(e-1)^{-1}\exp(-x).
  \end{eqnarray*}
Thus, $\bf U$ has finite moments of all orders. Subsequently, the existence of
$EW({\bf U})$ follows from that of $EZ({\bf U})$ by
\citet[lemma~A.2]{shi}.

Now, introduce $Z_d({\bf s})=\omega^{1-d}(1-\omega)^dW_d({\bf s})+\|{\bf s}\|^2$
and ${\bf U}_d=\arg\min_{\bf s}Z_d({\bf s})$, for $d=0,1$. Since
$EZ_d({\bf U}_d)$ exists and
$Z_d({\bf U}_d)\leq Z_d({\bf U})$, $EZ_d({\bf U})$ exists. Subsequently,
$EW_d({\bf U})$ exists. Similarly, $EW_d({\bf U}_{1-d})$ exists.
Since ${\bf U}_{1-d}$ is independent of process $W_d(\cdot)$,
$EW_d({\bf U}_{1-d})=0$. From $Z({\bf U})\leq Z({\bf U}_0)$,
we have
\[
\omega W_0({\bf U})+(1-\omega)W_1({\bf U})+\|{\bf U}\|^2
\leq \omega W_0({\bf U}_0)+(1-\omega)W_1({\bf U}_0)+\|{\bf U}_0\|^2.
\]
Upon taking expectations on both sides, we obtain
\[
(1-\omega)EW_1({\bf U})\leq E\{Z_0({\bf U}_0)-Z_0({\bf U})\}.
\]
Since $\Pr({\bf U}={\bf U}_0)=0$ as can be shown, $EW_1({\bf U})<0$.
Similarly, $EW_0({\bf U})<0$.

\subsubsection*{Proof of Corollary~\ref{coro2}}
For each class, the fold sizes differ by at most 1, which does not affect the
asymptotic results. Without loss of generality, we assume that both $n_0$ and
$n_1$ are divisible by $K$ so that the $K$ folds have equal sizes.

Write $\h{\phi}_d^{(k)}({\bf b}_{-\ell})$ and $\h{\phi}_d^{(-k)}({\bf b}_{-\ell})$
as the counterparts of $\h{\phi}_d({\bf b}_{-\ell})$ using data of the
$k$-th fold and data except the $k$-th fold, respectively, for
$k=1,\ldots,K$ and $d=0,1$. Extending Theorem~\ref{thm2} to address
subsets from the $K$-fold partition, we obtain
\begin{eqnarray*}
\lefteqn{\Big\{(n/K)^{2/3}[\h{\phi}_d^{(k)}\{\bt_{-\ell}+(n/K)^{-1/3}{\bf s}\}
    -\phi_d\{\bt_{-\ell}+(n/K)^{-1/3}{\bf s}\}
    -\h{\phi}_d^{(k)}(\bt_{-\ell})+\phi_d(\bt_{-\ell})],}\\
&& \hspace*{-.16in}\{(K-1)n/K\}^{2/3}\{\h{\phi}_d^{(-k)}[\bt_{-\ell}+\{(K-1)n/K\}^{-1/3}{\bf s}]\\
&& \hspace*{.2in}-\phi_d[\bt_{-\ell}+\{(K-1)n/K\}^{-1/3}{\bf s}]
   -\h{\phi}_d^{(-k)}(\bt_{-\ell})+\phi_d(\bt_{-\ell})\},\\
&& \hspace*{-.16in}n^{2/3}\{\h{\phi}_d(\bt_{-\ell}+n^{-1/3}{\bf s})-\phi_d(\bt_{-\ell}+n^{-1/3}{\bf s})
    -\h{\phi}_d(\bt_{-\ell})+\phi_d(\bt_{-\ell})\}
    :d=0,1;k=1,\ldots,K\Big\}\\
&& \rightsquigarrow
  \Big\{W_d^{(k)}({\bf s}),W_d^{(-k)}({\bf s}),W_d({\bf s}):d=0,1;k=1,\ldots,K\Big\},
\end{eqnarray*}
where Gaussian processes $W_d^{(k)}(\cdot)$, $d=0,1$, $k=1,\ldots,K$, are
independent of each other, $W_d^{(-k)}({\bf s})=(K-1)^{-1/3}
\sum_{l\neq k}W_d^{(l)}\{(K-1)^{-1/3}{\bf s}\}$, and
$W_d({\bf s})=K^{-1/3}\sum_{k=1}^KW_d^{(k)}(K^{-1/3}{\bf s})$. For each
$d=0,1$, $W_d^{(k)}(\cdot)$, $W_d^{(-k)}(\cdot)$, and $W_d(\cdot)$ have the same
distribution.
Let $W^{(-k)}({\bf s})=\omega W_0^{(-k)}({\bf s})+(1-\omega)W_1^{(-k)}({\bf s})$,
$Z^{(-k)}({\bf s})=W^{(-k)}({\bf s})+{\bf s}^\T{\bf H}{\bf s}/2$, and
${\bf U}^{(-k)}=\arg\min_{\bf s} Z^{(-k)}({\bf s})$.

Then, using the techniques in the proof of Theorem~\ref{thm3}, we can show
results~(\ref{cv1}) and~(\ref{cv2}) with
$V_d=K^{-1/3}\sum_{k=1}^K W_d^{(k)}\{(K-1)^{-1/3}{\bf U}^{(-k)}\}$ and
$V=\omega V_0+(1-\omega)V_1$.
Since $W_d^{(k)}(\cdot)$ is independent of ${\bf U}^{(-k)}$, $EV_d=EV=0$.
Results~(\ref{cvi}) and~(\ref{cvii}) follow from a generalized version of
the Delta method, with
\begin{eqnarray*}
  S_d &=& {\bf D}_d^\T
  [\{(K-1)/K\}^{-1/3}K^{-1}\sum_{k=1}^K{\bf U}^{(-k)}-{\bf U}],\\
  S &=& \{(K-1)/K\}^{-2/3}K^{-1}\sum_{k=1}^K{\bf U}^{(-k)\T}{\bf H}{\bf U}^{(-k)}/2
  -{\bf U}^\T{\bf HU}/2,
  \end{eqnarray*}
where ${\bf D}_d=\nabla \phi_d(\bt_{-\ell})$. Meanwhile, $S_d$ is non-degenerate as
\[
S_d=\{(K-1)/K\}^{-1/3}{\bf D}_d^\T
\left\{K^{-1}\sum_{k=1}^K\arg\min_{\bf s}Z^{(-k)}({\bf s})
-\arg\min_{\bf s}K^{-1}\sum_{k=1}^KZ^{(-k)}({\bf s})\right\}
\]
and $\arg\min$ is not a linear operator. The expectations of $S_d$ and $S$
follow from $E{\bf U}={\bf 0}$ and ${\bf U}^{(-k)}$ has the same distribution
as ${\bf U}$.

\subsubsection*{Proof of Proposition~\ref{prop1}}
Since $\h{\psi}^{(-k)}(\hb^{(-k)})\leq
\min_{{\bf b}:\|{\bf b}\|_1=1}\h{\psi}^{(-k)}({\bf b})+\epi_n$,
\begin{eqnarray*}
  (K-1)\sum_{k=1}^K\h{\psi}^{(-k)}(\hb^{(-k)})-K(K-1)\epi_n
  &\leq& \sum_{k=1}^K\sum_{l\neq k}\h{\psi}^{(-k)}(\hb^{(-l)})\\
  &=& \sum_{k=1}^K\sum_{l\neq k}\h{\psi}^{(-l)}(\hb^{(-k)})\\
  &=& \sum_{k=1}^K\h{\psi}^{(k)}(\hb^{(-k)})
  +\sum_{k=1}^K\frac{K-2}{K-1}\sum_{l\neq k}\h{\psi}^{(l)}(\hb^{(-k)})\\
  &=& \sum_{k=1}^K\left\{\h{\psi}^{(k)}(\hb^{(-k)})+(K-2)\h{\psi}^{(-k)}(\hb^{(-k)})\right\}.
\end{eqnarray*}
Thus,
\[
\sum_{k=1}^K\h{\psi}^{(-k)}(\hb^{(-k)})\leq\sum_{k=1}^K\h{\psi}^{(k)}(\hb^{(-k)})
+K(K-1)\epi_n,
\]
and the assertion follows.

\subsubsection*{Proof of Corollary~\ref{coro3}}
Following the proof of Corollary~\ref{coro2},
\[
n^{2/3}[\{\xbar{\psi}_{d,\mathrm{emp}}-\psi_{d,cv-ens}\}
  - \{\h{\psi}_d(\bt)-\psi_d(\bt)\}]
\rightsquigarrow \{(K-1)/K\}^{-2/3}K^{-1}\sum_{k=1}^KW_d^{(-k)}({\bf U}^{(-k)}),
\]
for $d=0,1$. It holds jointly with result~(\ref{cv1}), leading to
\[
n^{2/3}(\xbar{\psi}_{d,\mathrm{emp}}-\xbar{\psi}_{d,\mathrm{cv}})
\rightsquigarrow \{(K-1)/K\}^{-2/3}K^{-1}\sum_{k=1}^KW_d^{(-k)}({\bf U}^{(-k)})-V_d.\]
Since $Q(\cdot)$ is differentiable at $\psi_d(\bt)$,
\[
  Q(\xbar{\psi}_{d,\mathrm{emp}})-Q(\xbar{\psi}_{d,\mathrm{cv}})
  = Q'\{\psi_d(\bt)\}(\xbar{\psi}_{d,\mathrm{emp}}-\xbar{\psi}_{d,\mathrm{cv}})+o_p(n^{-2/3}).
\]
Similarly, $Q^{-1}(\cdot)$ is differentiable at $Q\{\psi_d(\bt)\}$ and we
have
\[
\h{\psi}_{d,\mathrm{cap}}=\h{\psi}_d(\hb)-\{(K-1)/K\}^{2/3}(\xbar{\psi}_{d,\mathrm{emp}}-\xbar{\psi}_{d,\mathrm{cv}})+o_p(n^{-2/3}).
\]
In light of result~(\ref{opt1}), we obtain
\[
n^{2/3}[\{\h{\psi}_{d,\mathrm{cap}}-\psi_d(\hb)\} -
  \{\h{\psi}_d(\bt)-\psi_d(\bt)\}]\rightsquigarrow
W_d({\bf U})
-K^{-1}\sum_{k=1}^KW_d^{(-k)}({\bf U}^{(-k)})
+\{(K-1)/K\}^{2/3} V_d.
\]
Accordingly, for the overall risk,
\[
n^{2/3}[\{\h{\psi}_{\mathrm{cap}}-\psi(\hb)\} -
  \{\h{\psi}(\bt)-\psi(\bt)\}]
\rightsquigarrow W({\bf U})
-K^{-1}\sum_{k=1}^KW^{(-k)}({\bf U}^{(-k)})
+\{(K-1)/K\}^{2/3} V.
\]
The assertions follow as $W_d^{(-k)}({\bf U}^{(-k)})$ and
$W^{(-k)}({\bf U}^{(-k)})$ have the same
distributions as $W_d({\bf U})$ and $W({\bf U})$, respectively.

\section*{Supplementary Materials}

\hspace*{1.2em}
Computational algorithm for linear ERM classification, additional simulation
results of Example~\ref{exp1}, additional simulation results of ERM
classification in Section~4.1, and simulation results of LR-ERM classification.

\section*{Funding}

This research was partially supported by the National Institutes
of Health grants R01 CA230268, R01 CA283687,
and P30 AI050409.

\newpage
\setcounter{page}{1}
\begin{center}
{\Large Supplementary Appendix}\hfill\phantom{nothing}\\[10pt]
\hfill{\Large\bf Cross-Audit Projection for Model Risk Prediction}\\
{\large Yijian {\sc Huang}}
\end{center}

\setcounter{section}{0}
\renewcommand{\thesection}{\Alph{section}}
\setcounter{equation}{0}
\renewcommand{\theequation}{S\arabic{equation}}
\setcounter{table}{0}
\renewcommand{\thetable}{S\arabic{table}}
\setcounter{figure}{0}
\renewcommand{\thefigure}{S\arabic{figure}}
\setcounter{algocf}{0}
\renewcommand{\thealgocf}{S\arabic{algocf}}

\section{Computational algorithm for linear ERM classification}

Upon replacing $I(x\leq0)$ with
$\sigma^{-1}\{(x-\sigma/2)^--(x+\sigma/2)^-\}$, problem~(\ref{est2}) becomes
\begin{eqnarray}
  \min_{{\bf b}:\|{\bf b}\|_1\leq1} && \sigma^{-1}\he\left\{
    \omega(-{\bf b}^\T{\bf X}_0^\circ-\sigma/2)^-+
    (1-\omega)({\bf b}^\T{\bf X}_1^\circ-\sigma/2)^-\right\}\nonumber\\
    && \mbox{}-\left[\sigma^{-1}\he\left\{
    \omega(-{\bf b}^\T{\bf X}_0^\circ+\sigma/2)^-+
    (1-\omega)({\bf b}^\T{\bf X}_1^\circ+\sigma/2)^-\right\}
    +c\|{\bf b}\|_1\right].\label{optim1}
\end{eqnarray}
\begin{algorithm}[!b]
\caption{Pseudo code for linear ERM classification}
\label{alg1}
  \DontPrintSemicolon
  \vspace*{.025in}
  \KwIn{initial value ${\bf b}_0$ of $\bf b$,
    starting value of $\sigma$,
    shrinking factor of $\sigma$
  }
  \KwOut{optimizer for $\bf b$}
  \vspace*{.05in}
      $i \longleftarrow 0$.\;
      \Repeat{sufficiently small improvement in objective function
        of problem~(\ref{est2}).}{
        \Repeat( \tcc{Concave-convex procedure with fixed $\sigma$: problem~(\ref{optim1})})
               {sufficiently small improvement in objective function
                 of problem~(\ref{optim1}).}{
                 1. {\em Convexify.} Form
                 \begin{eqnarray*}
                   T({\bf b},t;{\bf b}_i,t_i) &\equiv&
                   \sigma^{-1}\he\left\{
                   \omega(-{\bf b}_i^\T{\bf X}_0^\circ+\sigma/2)^-+
                  (1-\omega)({\bf b}_i^\T{\bf X}_1^\circ+\sigma/2)^-\right\}\\
                  &&\mbox{}+\sigma^{-1}\he\left\{
                   -\omega({\bf b}-{\bf b}_i)^\T{\bf X}_0^\circ I({\bf b}_i^\T{\bf X}_0^\circ\geq\sigma/2)\right.\\
                  && \mbox{}\left.\phantom{+\sigma^{-1}\he}+(1-\omega)({\bf b}-{\bf b}_i)^\T{\bf X}_1^\circ I({\bf b}_i^\T{\bf X}_1^\circ\leq-\sigma/2)\right\}\\
                  && \mbox{}+c\|{\bf b}_i\|_1+c\sign({\bf b}_i)^\T({\bf b}-{\bf b}_i).
                 \end{eqnarray*}
                 2. {\em Solve.} Set the value of ${\bf b}_{i+1}$ to the
                 solution of
                 \begin{eqnarray}
                  \min_{{\bf b}:\|{\bf b}\|_1\leq1} && \sigma^{-1}\he\left\{
                  \omega(-{\bf b}^\T{\bf X}_0^\circ-\sigma/2)^-+
                  (1-\omega)({\bf b}^\T{\bf X}_1^\circ-\sigma/2)^-\right\}\nonumber\\
                  && \mbox{}-T({\bf b},t;{\bf b}_i,t_i)\label{optim2}
                 \end{eqnarray}
                 3. {\em Update iteration.} $i \longleftarrow i+1$.\;
               }
               Shrink $\sigma$ value by the given factor.\;
      }
  \end{algorithm}
For a fixed $\sigma$, the concave-convex procedure \citep{yuille} is applied
for the optimization, which is the core of Algorithm~\ref{alg1}. At each
step, the concave component of the objective function is replaced with its
tangent plane at the current coefficient value resulting in a linear
program given by~(\ref{optim2}). 
Problem~(\ref{optim1}) is only an approximation of problem~(\ref{est2}).
A sequence
of decreasing $\sigma$ values are taken in the outer loop of
Algorithm~\ref{alg1}.

For the numerical studies in Section~4, the algorithm was implemented using a
linear programming solver from the {\tt R} package {\tt Rmosek}. The LR-ERM
classifier was
used for initialization. The initial value of $\sigma$ was set to the larger
of the two class-specific maximum gaps between adjacent order statistics of
the initial combinations. The shrinking factor for the $\sigma$ sequence was
set to 0.8, and the constant $c$ in problem~(\ref{est2}) was taken to be 2.

\section{Additional simulation results of Example~\ref{exp1}}

\begin{figure}[hb]
  \centerline{\includegraphics[width=6.5in]{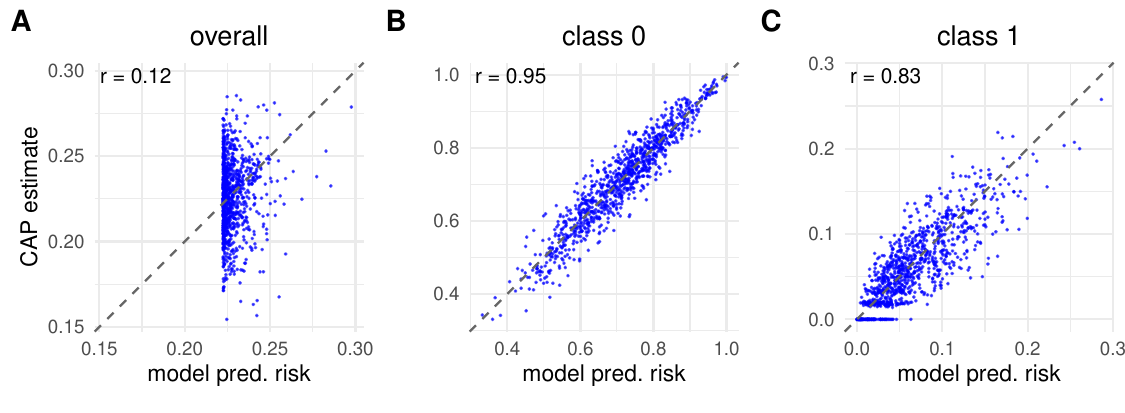}}
          {\caption{\label{figs0} Simulation results on CAP estimates versus
          model prediction risks in single-feature ERM
          binary classification with
          $X_0\sim N(0,1)$, $X_1\sim N(1,1)$, $\omega=0.25$, and
          $n_0=n_1=100$, based on 1000 simulated datasets.}}
\end{figure}

Figure~\ref{figs0} presents scatter plots of the CAP estimates against the
model prediction risks, along with the corresponding correlation coefficients.
For overall risk, the CAP estimator had a weak correlation with the model
prediction risk. In contrast, for class-specific risks, strong correlations
were observed demonstrating that the CAP estimators
tracked the model prediction risks more closely than the algorithm prediction
risks; see Remark~\ref{rmk1}.

\section{Additional simulation results of ERM classification in Section~4.1}

\begin{figure}[bt]
  \centerline{\includegraphics[width=5.8in]{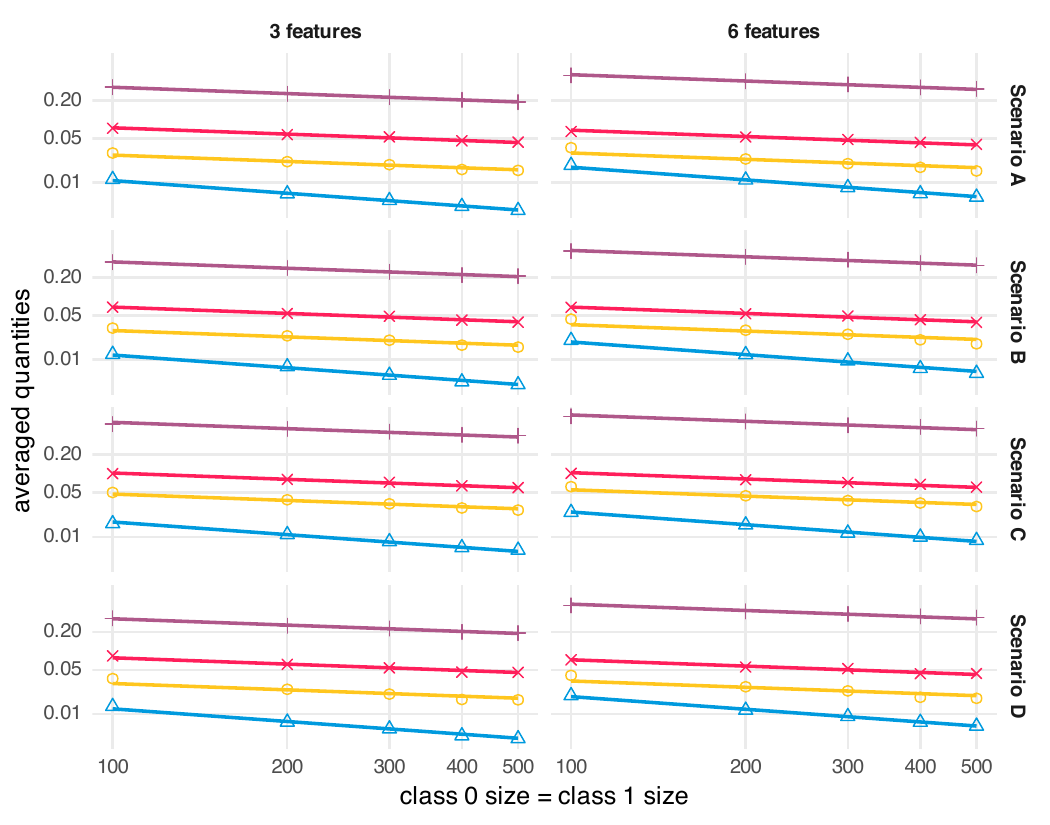}}
  {\caption{\label{figs1}Simulation results on
  $\|\hb-\bt\|_1$ ($+$), $\psi(\bt)-\psi(\hb)$ ({\tiny $\triangle$}),
  $|\psi_0(\bt)-\psi_0(\hb)|$ ($\times$), and
  $|\psi_1(\bt)-\psi_1(\hb)|$ ($\circ$). Each straight line is the
  least-squares fitting of an averaged quantity versus sample size, both on the
  logarithmic scale, with the slope fixed to $-2/3$ for $\psi(\bt)-\psi(\hb)$
  and to $-1/3$ for others.}}
\end{figure}

Figure~\ref{figs1} shows mean absolute error of $\hb$ and mean absolute
differences between model prediction risks and the risks of the optimal
classifier.
Across all set-ups, mean $\psi(\bt)-\psi(\hb)$ diminished with the sample
size at an approximate rate of $n^{-2/3}$, whereas the
others exhibited a rate of about $n^{-1/3}$. These findings corroborate the
asymptotic results of cube-root convergence of $\hb$.

\begin{table}[t!]
  \caption{\label{tabs1}Simulation results on overall $|$ class~0 $|$ class~1
    risk prediction of the estimated classifier with three features}
\centering
\begin{tabular}{
    c@{\hspace*{8pt}}
    r@{$|$}r@{$|$}r@{\hspace*{6pt}}r@{$|$}r@{$|$}r@{\hspace*{8pt}}
    r@{$|$}r@{$|$}r@{\hspace*{6pt}}r@{$|$}r@{$|$}r@{\hspace*{8pt}}
    r@{$|$}r@{$|$}r@{\hspace*{6pt}}r@{$|$}r@{$|$}r@{\hspace*{6pt}}r@{$|$}r@{$|$}r}\\\hline\hline\\[-10pt]
  $n_0\phantom{=}$ & \multicolumn{6}{c}{rep.\ 2-fold CV} & \multicolumn{6}{c}{rep.\ 10-fold CV}
  & \multicolumn{9}{c}{rep.\ 10-fold CAP}\\
  \cmidrule(lr){2-7}\cmidrule(lr){8-13}\cmidrule(lr){14-22}\\[-10pt]
  $\phantom{,}=n_1$ &
  \multicolumn{3}{c}{B} & \multicolumn{3}{c}{D} &
  \multicolumn{3}{c}{B} & \multicolumn{3}{c}{D} &
  \multicolumn{3}{c}{B} & \multicolumn{3}{c}{D} & \multicolumn{3}{c}{C}
  \\\hline\\[-10pt]
  \multicolumn{22}{c}{Scenario A}\\
100 &  9 &$-$10 & 15 &20 &59 &32 & 3 &$-$3 & 4 &21 &49 &27 & 0 &$-$2 & 1 &21 &40 &24 &94.2 &94.1 &94.8\\
200 & 4 &$-$7 & 8 &14 &44 &22 &1 &0 &1 &14 &36 &19 &0 &0 &0 &15 &27 &17 &93.5 &95.1 &92.9\\
300 & 3 &$-$8 & 7 &11 &38 &18 &0 &0 &0 &11 &31 &15 &$-$1 &$-$1 &$-$1 &12 &23 &13 &94.4 &94.6 &95.2\\
400 & 2 &$-$5 & 5 &10 &34 &15 & 0 &$-$2 & 1 &10 &27 &13 &$-$1 &$-$1 & 0 &10 &20 &11 &93.8 &93.5 &94.7\\
500 & 2 &$-$2 & 3 & 9 &30 &13 &0 &1 &0 & 9 &24 &11 &0 &0 &0 & 9 &18 &10 &93.7 &95.5 &93.7\\
  \multicolumn{22}{c}{Scenario B}\\
100 & 10 &$-$12 & 17 &21 &56 &32 & 4 &$-$1 & 5 &21 &48 &26 &1 &0 &1 &22 &40 &24 &92.0 &95.0 &93.3\\
200 & 5 &$-$7 & 9 &14 &40 &21 &1 &1 &1 &14 &35 &18 & 0 & 1 &$-$1 &15 &29 &17 &92.6 &95.4 &92.3\\
300 &  4 &$-$11 &  9 &11 &35 &18 & 0 &$-$1 & 1 &11 &29 &14 & 0 &$-$1 & 0 &11 &23 &12 &94.1 &94.7 &94.2\\
400 & 3 &$-$8 & 6 & 9 &31 &14 & 0 &$-$3 & 1 & 9 &25 &12 & 0 &$-$2 & 0 &10 &19 &10 &93.8 &94.5 &95.6\\
500 & 2 &$-$6 & 5 & 8 &29 &13 & 0 &$-$1 & 1 & 9 &24 &11 & 0 &$-$1 & 0 & 9 &19 & 9 &93.0 &94.1 &94.7\\
  \multicolumn{22}{c}{Scenario C}\\
100 & 7 &$-$9 &13 &20 &60 &33 & 0 &$-$2 & 1 &21 &51 &28 &$-$2 &$-$4 &$-$1 &22 &41 &25 &93.2 &93.9 &93.9\\
200 & 4 &$-$9 & 9 &15 &46 &23 &$-$1 &$-$2 & 0 &15 &38 &19 &$-$2 &$-$2 &$-$2 &15 &29 &16 &92.9 &93.1 &92.9\\
300 &  4 &$-$14 & 11 &12 &39 &20 & 0 &$-$2 & 1 &12 &32 &15 &$-$1 & 0 &$-$1 &12 &24 &13 &94.0 &94.9 &95.2\\
400 &  4 &$-$12 &  9 &10 &36 &18 & 1 &$-$1 & 1 &10 &28 &14 &0 &1 &0 &10 &20 &11 &94.8 &94.9 &94.6\\
500 &  3 &$-$15 &  9 & 9 &36 &18 & 0 &$-$3 & 1 & 9 &27 &13 &$-$1 & 0 &$-$1 & 9 &18 &10 &93.7 &94.4 &92.9\\
  \multicolumn{22}{c}{Scenario D}\\
100 & 9 &$-$9 &15 &23 &62 &36 &2 &0 &2 &24 &49 &31 &$-$1 & 1 &$-$2 &24 &39 &28 &93.1 &95.5 &92.7\\
200 & 5 &$-$9 &10 &15 &46 &25 & 1 &$-$2 & 2 &15 &35 &20 &$-$1 & 0 &$-$1 &15 &27 &17 &93.4 &94.8 &94.4\\
300 & 4 &$-$9 & 8 &12 &39 &19 &0 &0 &0 &12 &31 &16 & 0 &$-$1 & 0 &12 &23 &14 &93.8 &94.7 &93.9\\
400 & 3 &$-$7 & 6 & 9 &34 &16 & 0 &$-$1 & 0 &10 &27 &13 &$-$1 & 0 &$-$1 &10 &20 &11 &93.7 &94.6 &94.1\\
500 & 2 &$-$7 & 5 & 9 &33 &15 & 0 &$-$2 & 0 & 9 &26 &12 &$-$1 & 0 &$-$1 & 9 &18 &10 &94.0 &95.0 &94.4\\
\hline\\[-10pt]
\multicolumn{22}{l}{\footnotesize
  B: bias ($\times1000$), D: mean absolute
  deviation ($\times1000$), C: coverage probability of 95\% risk}\\
\multicolumn{22}{l}{\footnotesize
  bound (\%).}
\end{tabular}
\end{table}

Tables~\ref{tabs1} and~\ref{tabs2} present simulation results for additional
model risk prediction methods, repeated 2-fold CV,
repeated 10-fold CV, and repeated 10-fold CAP, all with 16 repetitions,
supplementing Tables~\ref{tab1} and~\ref{tab2},
respectively. They
provide (i)~comparisons between CV and CAP with the same fold number 
$K$ and the same number of repetitions: repeated 2-fold CV versus repeated
2-fold CAP, and repeated 10-fold CV versus repeated 10-fold CAP; and (ii)~a
comparison between repeated 2-fold CAP and repeated 10-fold CAP. $K$-fold
CV was sensitive to the choice of $K$, exhibiting larger positive bias
for smaller values of $K$. In contrast, $K$-fold CAP showed little performance
difference for different choices of $K$. These results
further confirm that CAP outperforms CV, particularly
for class-specific risks.

\begin{table}[t!]
  \caption{\label{tabs2}Simulation results on overall $|$ class~0 $|$ class~1
    risk prediction of the estimated classifier with six features}
\centering
\begin{tabular}{
    c@{\hspace*{8pt}}
    r@{$|$}r@{$|$}r@{\hspace*{6pt}}r@{$|$}r@{$|$}r@{\hspace*{8pt}}
    r@{$|$}r@{$|$}r@{\hspace*{6pt}}r@{$|$}r@{$|$}r@{\hspace*{8pt}}
    r@{$|$}r@{$|$}r@{\hspace*{6pt}}r@{$|$}r@{$|$}r@{\hspace*{6pt}}r@{$|$}r@{$|$}r}\\\hline\hline\\[-10pt]
  $n_0\phantom{=}$ & \multicolumn{6}{c}{rep.\ 2-fold CV} & \multicolumn{6}{c}{rep.\ 10-fold CV}
  & \multicolumn{9}{c}{rep.\ 10-fold CAP}\\
  \cmidrule(lr){2-7}\cmidrule(lr){8-13}\cmidrule(lr){14-22}\\[-10pt]
  $\phantom{,}=n_1$ &
  \multicolumn{3}{c}{B} & \multicolumn{3}{c}{D} &
  \multicolumn{3}{c}{B} & \multicolumn{3}{c}{D} &
  \multicolumn{3}{c}{B} & \multicolumn{3}{c}{D} & \multicolumn{3}{c}{C}
  \\\hline\\[-10pt]
  \multicolumn{22}{c}{Scenario A}\\
100 & 13 &$-$14 & 22 &23 &58 &36 &2 &0 &2 &22 &50 &28 &$-$2 & 0 &$-$2 &24 &42 &28 &91.2 &93.4 &91.5\\
200 & 7 &$-$9 &12 &15 &43 &23 & 0 &$-$2 & 1 &15 &36 &18 &$-$2 &$-$1 &$-$2 &15 &29 &17 &92.5 &93.1 &94.0\\
300 &  5 &$-$12 & 10 &12 &36 &20 & 0 &$-$2 & 1 &12 &29 &16 &$-$1 & 0 &$-$2 &13 &24 &14 &92.7 &94.1 &93.3\\
400 & 4 &$-$6 & 7 &10 &33 &17 &0 &0 &0 &10 &27 &14 &$-$1 &$-$1 &$-$1 &11 &20 &12 &92.5 &93.9 &92.9\\
500 & 5 &$-$2 & 7 & 9 &31 &14 &1 &2 &1 & 9 &25 &12 &1 &2 &1 & 9 &17 &10 &94.9 &95.5 &95.8\\
  \multicolumn{22}{c}{Scenario B}\\
100 & 14 &$-$17 & 25 &24 &53 &38 & 2 &$-$1 & 3 &22 &48 &28 &$-$3 & 0 &$-$3 &25 &42 &28 &88.4 &94.1 &88.5\\
200 &  7 &$-$11 & 13 &15 &42 &23 & 0 &$-$3 & 1 &14 &36 &18 &$-$2 &$-$1 &$-$2 &15 &29 &17 &91.1 &93.6 &93.7\\
300 &  5 &$-$11 & 11 &12 &35 &19 & 0 & 1 &$-$1 &11 &30 &15 &$-$1 & 1 &$-$2 &11 &24 &13 &92.7 &94.3 &92.8\\
400 & 5 &$-$9 & 9 &10 &31 &16 & 0 &$-$2 & 1 &10 &26 &13 &$-$1 & 0 &$-$1 &10 &20 &11 &93.9 &95.4 &93.1\\
500 & 5 &$-$5 & 8 & 9 &28 &14 &1 &1 &1 & 9 &23 &11 &0 &1 &0 & 9 &17 &10 &93.6 &95.6 &94.9\\
  \multicolumn{22}{c}{Scenario C}\\
100 & 12 &$-$15 & 21 &24 &58 &39 & 1 &$-$2 & 1 &22 &49 &29 &$-$4 &$-$3 &$-$4 &24 &40 &28 &91.6 &94.8 &91.2\\
200 &  8 &$-$14 & 15 &16 &44 &27 & 0 &$-$2 & 1 &16 &37 &20 &$-$2 & 0 &$-$2 &16 &29 &18 &92.5 &94.2 &92.7\\
300 &  7 &$-$17 & 15 &13 &39 &23 & 1 &$-$2 & 1 &12 &31 &16 &$-$1 & 0 &$-$1 &12 &24 &14 &93.3 &93.6 &93.7\\
400 &  5 &$-$11 & 11 &10 &34 &18 & 1 &$-$1 & 2 &10 &27 &14 & 0 & 1 &$-$1 &10 &21 &11 &94.3 &94.7 &94.9\\
500 &  5 &$-$13 & 10 & 9 &33 &17 & 1 &$-$3 & 2 & 9 &26 &13 &$-$1 & 0 &$-$1 & 9 &18 &10 &94.9 &94.8 &93.8\\
  \multicolumn{22}{c}{Scenario D}\\
100 & 15 &$-$15 & 25 &25 &59 &40 &3 &0 &4 &22 &50 &30 &$-$1 & 2 &$-$2 &24 &40 &29 &91.3 &95.2 &91.2\\
200 & 8 &$-$9 &14 &16 &42 &25 &1 &1 &2 &15 &35 &20 & 0 & 3 &$-$1 &16 &28 &18 &93.1 &95.8 &92.9\\
300 &  5 &$-$15 & 12 &12 &41 &22 & 0 &$-$2 & 1 &12 &31 &17 &$-$1 & 0 &$-$1 &12 &22 &14 &92.7 &94.2 &95.3\\
400 & 5 &$-$9 & 9 &10 &35 &18 & 0 &$-$2 & 1 &11 &28 &14 &$-$1 & 0 &$-$1 &11 &21 &12 &92.8 &94.3 &93.5\\
500 & 5 &$-$6 & 8 & 9 &32 &16 &1 &0 &1 & 9 &25 &13 &0 &0 &0 &10 &18 &11 &94.6 &93.5 &95.2\\
\hline\\[-10pt]
\multicolumn{22}{l}{\footnotesize
  B: bias ($\times1000$), D: mean absolute
  deviation ($\times1000$), C: coverage probability of 95\% risk}\\
\multicolumn{22}{l}{\footnotesize
  bound (\%).}
\end{tabular}
\end{table}

\section{Simulation results of LR-ERM classification}

LR-ERM classification is commonly used in practice when multiple features are
present. It first employs logistic regression to combine features, followed
by ERM to estimate the classification threshold. As noted in
Remark~\ref{remarka}, the resulting combination can be treated asymptotically
as a fixed single feature for the purpose of model risk prediction.
Under the same set-ups in Section~4.1,
Tables~\ref{tabs3} and~\ref{tabs4} present simulation results for model risk
prediction with LR-ERM classification. The bootstrap method is that of
\citet{harrell} with 100 resamples, and repeat 2-fold CAP has 16 repetitions.
The observed patterns were similar to
those reported in Tables~\ref{tab1} and~\ref{tab2} with ERM classification.

\begin{sidewaystable}[htbp]
  \caption{\label{tabs3}Simulation results on overall $|$ class~0 $|$ class~1
    risk prediction of the LR-ERM classifier with three features}
\centering
\begin{tabular}{
    c@{\hspace*{8pt}}
    r@{$|$}r@{$|$}r@{\hspace*{6pt}}r@{$|$}r@{$|$}r@{\hspace*{8pt}}
    r@{$|$}r@{$|$}r@{\hspace*{6pt}}r@{$|$}r@{$|$}r@{\hspace*{8pt}}
    r@{$|$}r@{$|$}r@{\hspace*{6pt}}r@{$|$}r@{$|$}r@{\hspace*{8pt}}
    r@{$|$}r@{$|$}r@{\hspace*{6pt}}r@{$|$}r@{$|$}r@{\hspace*{6pt}}r@{$|$}r@{$|$}r}\\\hline\hline\\[-10pt]
  $n_0\phantom{=}$ & \multicolumn{6}{c}{empirical} & \multicolumn{6}{c}{bootstrap} & \multicolumn{6}{c}{10-fold CV}
  & \multicolumn{9}{c}{repeated 2-fold CAP}\\
  \cmidrule(lr){2-7}\cmidrule(lr){8-13}\cmidrule(lr){14-19}\cmidrule(lr){20-28}\\[-10pt]
  $\phantom{,}=n_1$ &
  \multicolumn{3}{c}{B} & \multicolumn{3}{c}{D} & \multicolumn{3}{c}{B} &
  \multicolumn{3}{c}{D} & \multicolumn{3}{c}{B} & \multicolumn{3}{c}{D} &
  \multicolumn{3}{c}{B} & \multicolumn{3}{c}{D} & \multicolumn{3}{c}{C}
  \\\hline\\[-10pt]
  \multicolumn{28}{c}{Scenario A}\\
100 &-27 &-26 &-26 &29 &44 &29 &-8 &-8 &-8 &19 &39 &21 &-1 &-2 &-1 &21 &55 &28 &-3 &-3 &-2 &19 &39 &22 &93.8 &94.4 &93.7\\
200 &-16 &-14 &-16 &18 &30 &19 &-4 &-3 &-4 &13 &28 &15 & 1 &-1 & 1 &14 &41 &20 & 0 & 1 &-1 &13 &28 &15 &95.2 &95.1 &95.5\\
300 &-12 &-11 &-12 &14 &24 &15 &-3 &-2 &-3 &10 &22 &12 &0 &0 &0 &11 &34 &15 & 0 & 1 &-1 &10 &22 &12 &95.4 &95.8 &95.0\\
400 &-10 &-10 &-10 &12 &21 &12 &-2 &-2 &-2 & 9 &20 &10 & 0 &-2 & 1 &10 &32 &14 &0 &0 &0 & 9 &20 &10 &94.8 &95.4 &95.1\\
500 &-9 &-8 &-9 &11 &19 &11 &-2 &-2 &-3 & 8 &18 & 9 &0 &0 &0 & 9 &28 &12 &-1 & 0 &-1 & 8 &18 & 9 &94.5 &94.5 &95.2\\
  \multicolumn{28}{c}{Scenario B}\\
100 &-27 &-24 &-26 &28 &43 &29 &-8 &-6 &-8 &19 &39 &21 &-1 & 0 &-1 &21 &53 &27 &-3 &-1 &-3 &19 &39 &22 &93.5 &94.6 &94.1\\
200 &-16 &-14 &-16 &17 &30 &18 &-4 &-3 &-4 &12 &28 &14 &0 &2 &0 &14 &42 &19 & 0 & 1 &-1 &13 &28 &15 &94.5 &95.4 &94.4\\
300 &-12 &-12 &-12 &14 &23 &14 &-3 &-3 &-3 &10 &22 &11 &0 &0 &0 &11 &34 &16 &-1 & 0 &-1 &10 &22 &12 &95.5 &95.0 &95.9\\
400 &-10 &-10 &-10 &12 &21 &12 &-2 &-3 &-2 & 9 &20 &10 &0 &0 &0 &10 &33 &14 & 0 &-1 & 0 & 9 &20 &10 &94.8 &94.1 &95.2\\
500 &-9 &-9 &-9 &11 &19 &11 &-2 &-3 &-2 & 8 &18 & 9 & 0 & 1 &-1 & 9 &28 &13 &-1 &-1 &-1 & 8 &18 & 9 &93.7 &94.2 &94.5\\
  \multicolumn{28}{c}{Scenario C}\\
100 &-28 &-28 &-27 &30 &44 &30 &-8 &-7 &-8 &20 &39 &23 & 0 &-3 & 1 &23 &60 &32 &-2 &-2 &-2 &21 &39 &24 &93.4 &94.5 &93.3\\
200 &-18 &-17 &-17 &19 &32 &20 &-4 &-4 &-5 &14 &29 &16 &0 &2 &0 &16 &47 &22 &0 &0 &0 &14 &29 &16 &95.7 &93.7 &95.8\\
300 &-14 &-16 &-13 &16 &26 &17 &-4 &-6 &-3 &12 &23 &14 &-1 & 0 &-1 &13 &40 &18 &-1 &-3 &-1 &12 &22 &14 &94.2 &93.8 &95.6\\
400 &-11 &-12 &-10 &13 &21 &13 &-2 &-3 &-2 &10 &19 &11 &0 &1 &0 &11 &34 &16 & 0 &-1 & 0 &10 &19 &12 &95.3 &96.1 &95.6\\
500 & -9 &-10 & -9 &11 &19 &12 &-2 &-2 &-2 & 9 &18 &10 &0 &1 &0 &10 &31 &14 &0 &0 &0 & 9 &18 &10 &95.0 &94.8 &95.8\\
  \multicolumn{28}{c}{Scenario D}\\
100 &-28 &-23 &-29 &30 &41 &31 &-7 &-5 &-8 &20 &37 &23 &0 &0 &0 &22 &54 &30 &-2 & 1 &-2 &20 &37 &23 &95.2 &95.2 &94.0\\
200 &-18 &-15 &-18 &19 &30 &21 &-5 &-4 &-5 &14 &28 &16 &-1 & 1 &-1 &15 &44 &21 &-1 &-1 &-1 &14 &28 &16 &95.0 &94.0 &94.9\\
300 &-13 &-12 &-12 &15 &25 &15 &-3 &-4 &-3 &11 &23 &12 & 1 &-2 & 1 &12 &38 &17 & 0 &-1 & 0 &11 &23 &12 &96.0 &95.2 &96.1\\
400 &-11 &-10 &-11 &13 &22 &13 &-3 &-3 &-2 &10 &20 &11 &0 &0 &0 &11 &33 &15 & 0 &-1 & 0 &10 &20 &11 &96.1 &94.0 &96.6\\
500 & -9 &-10 & -9 &11 &19 &12 &-2 &-3 &-2 & 9 &18 &10 &0 &0 &0 & 9 &31 &14 & 0 &-1 & 0 & 9 &18 &10 &94.7 &93.3 &95.4\\
\hline\\[-10pt]
\multicolumn{28}{l}{\footnotesize
  B: bias ($\times1000$), D: mean absolute
  deviation ($\times1000$), C: coverage probability of 95\% risk bound (\%).}
\end{tabular}
\end{sidewaystable}

\begin{sidewaystable}[htbp]
  \caption{\label{tabs4}Simulation results on overall $|$ class~0 $|$ class~1
    risk prediction of the LR-ERM classifier with six features}
\centering
\begin{tabular}{
    c@{\hspace*{8pt}}
    r@{$|$}r@{$|$}r@{\hspace*{6pt}}r@{$|$}r@{$|$}r@{\hspace*{8pt}}
    r@{$|$}r@{$|$}r@{\hspace*{6pt}}r@{$|$}r@{$|$}r@{\hspace*{8pt}}
    r@{$|$}r@{$|$}r@{\hspace*{6pt}}r@{$|$}r@{$|$}r@{\hspace*{8pt}}
    r@{$|$}r@{$|$}r@{\hspace*{6pt}}r@{$|$}r@{$|$}r@{\hspace*{6pt}}r@{$|$}r@{$|$}r}\\\hline\hline\\[-10pt]
  $n_0\phantom{=}$ & \multicolumn{6}{c}{empirical} & \multicolumn{6}{c}{bootstrap} & \multicolumn{6}{c}{10-fold CV}
  & \multicolumn{9}{c}{repeated 2-fold CAP}\\
  \cmidrule(lr){2-7}\cmidrule(lr){8-13}\cmidrule(lr){14-19}\cmidrule(lr){20-28}\\[-10pt]
  $\phantom{,}=n_1$ &
  \multicolumn{3}{c}{B} & \multicolumn{3}{c}{D} & \multicolumn{3}{c}{B} &
  \multicolumn{3}{c}{D} & \multicolumn{3}{c}{B} & \multicolumn{3}{c}{D} &
  \multicolumn{3}{c}{B} & \multicolumn{3}{c}{D} & \multicolumn{3}{c}{C}
  \\\hline\\[-10pt]
  \multicolumn{28}{c}{Scenario A}\\
100 &-32 &-29 &-31 &32 &45 &32 &-6 &-5 &-7 &19 &38 &21 &2 &1 &2 &22 &57 &30 &1 &2 &1 &20 &39 &23 &94.8 &96.0 &93.7\\
200 &-19 &-18 &-19 &20 &31 &21 &-4 &-4 &-4 &13 &28 &15 &0 &0 &0 &14 &42 &20 &0 &0 &0 &13 &28 &15 &95.0 &95.0 &94.3\\
300 &-14 &-13 &-14 &16 &25 &16 &-3 &-2 &-3 &11 &23 &12 &0 &1 &0 &12 &38 &17 &0 &1 &0 &11 &23 &12 &95.7 &95.7 &94.8\\
400 &-11 &-12 &-11 &13 &22 &13 &-2 &-3 &-2 & 9 &20 &10 &1 &0 &1 &10 &34 &15 &0 &0 &1 & 9 &20 &11 &96.8 &95.3 &95.9\\
500 &-10 &-10 & -9 &11 &20 &11 &-2 &-3 &-1 & 8 &19 & 9 &1 &0 &1 & 9 &30 &13 & 1 &-1 & 1 & 8 &18 & 9 &94.9 &94.3 &95.8\\
  \multicolumn{28}{c}{Scenario B}\\
100 &-31 &-28 &-30 &32 &44 &32 &-5 &-3 &-6 &18 &38 &20 &3 &0 &4 &20 &57 &28 &1 &3 &1 &19 &38 &22 &94.9 &95.4 &94.8\\
200 &-19 &-18 &-19 &21 &31 &21 &-4 &-4 &-5 &13 &28 &15 & 0 &-1 & 0 &14 &42 &20 &0 &0 &0 &13 &28 &15 &95.7 &95.1 &94.6\\
300 &-15 &-14 &-14 &16 &25 &16 &-3 &-3 &-3 &10 &23 &12 & 0 &-1 & 0 &11 &37 &16 & 0 & 0 &-1 &10 &23 &12 &95.7 &94.0 &94.9\\
400 &-11 &-11 &-11 &13 &22 &13 &-2 &-2 &-2 & 9 &20 &10 &1 &0 &1 &10 &33 &15 &0 &1 &0 & 9 &20 &10 &95.4 &94.2 &94.9\\
500 &-10 &-11 & -9 &11 &20 &11 &-2 &-3 &-1 & 8 &18 & 9 & 1 &-3 & 2 & 9 &30 &13 & 0 &-1 & 1 & 8 &18 & 9 &95.9 &94.0 &95.8\\
  \multicolumn{28}{c}{Scenario C}\\
100 &-35 &-36 &-33 &35 &47 &35 &-8 &-8 &-8 &21 &38 &23 & 1 &-5 & 2 &23 &61 &33 & 0 &-1 & 0 &21 &38 &25 &95.3 &95.1 &95.0\\
200 &-22 &-21 &-21 &22 &32 &23 &-5 &-4 &-5 &14 &29 &17 &-1 &-1 & 0 &15 &47 &23 &-1 & 0 &-1 &14 &28 &16 &95.0 &94.5 &95.7\\
300 &-15 &-13 &-15 &16 &24 &17 &-2 &-1 &-3 &11 &22 &13 &2 &0 &2 &13 &43 &20 &1 &3 &0 &11 &22 &13 &95.9 &95.6 &95.1\\
400 &-13 &-13 &-13 &15 &22 &15 &-3 &-3 &-3 &10 &20 &12 &0 &0 &0 &11 &38 &17 &0 &0 &0 &10 &20 &12 &95.2 &94.6 &95.1\\
500 &-11 &-11 &-11 &12 &19 &13 &-2 &-2 &-2 & 9 &18 &10 &1 &1 &1 &10 &34 &14 &0 &0 &0 & 9 &18 &10 &94.7 &95.3 &95.1\\
  \multicolumn{28}{c}{Scenario D}\\
100 &-35 &-32 &-35 &36 &45 &37 &-8 &-8 &-9 &21 &38 &23 & 0 &-4 & 1 &22 &57 &31 &-1 &-1 & 0 &21 &37 &25 &94.0 &95.3 &93.7\\
200 &-20 &-18 &-20 &21 &31 &22 &-4 &-3 &-4 &13 &28 &15 & 1 &-2 & 2 &15 &46 &20 &1 &1 &1 &13 &29 &15 &95.8 &94.8 &95.7\\
300 &-15 &-14 &-15 &16 &26 &17 &-3 &-2 &-3 &11 &23 &13 &1 &3 &0 &13 &40 &18 &1 &1 &1 &11 &23 &13 &97.0 &94.5 &96.6\\
400 &-13 &-11 &-13 &14 &22 &15 &-3 &-2 &-3 & 9 &21 &11 & 0 &-1 & 0 &10 &36 &16 &0 &0 &0 & 9 &21 &11 &96.0 &94.8 &95.9\\
500 &-10 &-10 &-10 &12 &20 &12 &-2 &-2 &-2 & 9 &18 &10 & 0 &-2 & 1 &10 &31 &14 &1 &0 &1 & 9 &18 &10 &95.9 &94.4 &96.1\\
\hline\\[-10pt]
\multicolumn{28}{l}{\footnotesize
  B: bias ($\times1000$), D: mean absolute
  deviation ($\times1000$), C: coverage probability of 95\% risk bound (\%).}
\end{tabular}
\end{sidewaystable}

\end{document}